\documentclass[10pt]{article}
\usepackage{amsmath, amssymb,amsthm}
\usepackage{amsfonts, graphics, graphicx, latexsym, epsfig}
\usepackage{cleveref}
\usepackage{color}
\usepackage{dsfont}
\usepackage{enumerate}
\newenvironment{rcases}
  {\left.\begin{aligned}}
  {\end{aligned}\right\rbrace}

\numberwithin{equation}{section}
\newtheorem{theo}{{\bf{Theorem}}}[section]
\newtheorem{prop}[theo]{{\bf Proposition}}
\newtheorem{lem}[theo]{{\bf Lemma}}

\newtheorem{coro}[theo]{{\bf Corollary}}

\newtheorem{rem}{{\bf Remark}}[section]

\newtheorem{defi}{{\bf Definition}}[section]

\renewcommand{\proof}{\noindent{\bf Proof.\ }}

\newcommand{\no}{\nonumber}
\newcommand{\noi}{\noindent}

\newcommand{\pa}{\partial}

\newcommand{\RR}{\mathbb{R}}
\newcommand{\D}{\mathcal{D}}

\newcommand{\la}{\lambda}

\newcommand{\calF}{\mathcal{F}}
\newcommand{\X}{\mathcal{X}}

\newcommand{\tab}{\hspace*{0.3in}}

\newcommand{\vf}{\varphi}

\newcommand{\vp}{\varepsilon}

\catcode`\@=11
\catcode`\@=12

\begin{document}

\title{Optimal Hedging in a Regime Switching Jump Diffusion Model \thanks{This research was supported in part by the SERB MATRICS (MTR/2017/000543) and DST FIST (SR/FST/MSI-105).}}
\author{Anindya Goswami\thanks{Correponding author: IISER, Pune 411008, India; email: anindya@iiserpune.ac.in.}\qquad\qquad Omkar Manjarekar\thanks{IISER, Pune 411008, India; email: omkar.m26@gmail.com.}\qquad\qquad Anjana R\thanks{Department of Mathematics, University of Botswana, Gaborone, Botswana; email: anjana\_rama@hotmail.com.} }

\date{}

\maketitle

\begin{abstract}
\noi This paper presents the solution to a European option pricing problem by considering a regime-switching jump diffusion model for the underlying financial asset price dynamics. The regimes are assumed to be the results of an observed pure jump process, driving the values of interest rate and volatility coefficient. The pure jump process is assumed to be a semi-Markov process on finite state space. This consideration helps to incorporate a specific type of memory influence in the asset price. Under this model assumption, the locally risk minimizing prices of the European type path-independent options are found. The F\"{o}llmer-Schweizer decomposition is adopted to show that the option price satisfies an evolution problem, as a function of time, stock price, market regime, and the stagnancy period. To be more precise, the evolution problem involves a linear, parabolic, degenerate and non-local system of integro-partial differential equations. We have established existence and uniqueness of the classical solution to the evolution problem in an appropriate class. This has helped us in obtaining the optimal hedging.
\end{abstract}

\noi{\bf Keywords:} jump-diffusion model, semi-Markov processes, locally risk minimizing pricing, optimal hedging, generalized solution, integro partial differential equation.

\noi {\bf Classification No:} 60K15, 91B30, 91G20, 91G60.

\section{Introduction}
There are extensive literature available in the theory and practice of option valuation following the pioneering work by Black and Scholes \cite{BS} in 1973. Contrary to the subsequent empirical evidence from the dynamics of financial asset prices, the Black-Scholes-Merton (BSM) model assumed a constant growth rate and a constant or deterministic volatility coefficient. The classical option pricing theory of BSM also relies on complete markets in which the payoff of every contingent claim can be replicated by a self-financing portfolio consisting of investments in the underlying stock and in a money market account paying a risk-less rate of interest. Hence an investor can be completely hedged against the risk of writing an option.

In subsequent studies, to overcome the limitations of BSM model, various option valuation models have been proposed and implemented in tune with increasingly more realistic price dynamics. These include stochastic volatility models, jump-diffusion models, regime-switching models and combinations of these. The markets in these models are incomplete where a perfect hedge is impossible. Various approaches have been adopted in solving option pricing in incomplete markets. The \emph{local risk minimization approach} is one such and was initiated by F\"{o}llmer, Sonderman, and Schweizer  \cite{FSond, Sch1990, Sch1992, Sch1993, Sch2001}. To hedge a claim in this approach, a unique dynamic strategy is sought for that replicates the claim at the maturity by allowing additional cash flow with continuous trading. This unique strategy minimizes the quadratic residual risk (QRR), a measure of the accumulated additional cash flow, under a certain set of constraints. This minimizing strategy is termed as the \textit{optimal hedging}. Thus in a complete market, a self financing hedging strategy becomes the optimal hedging and results in zero QRR. The existence of an optimal hedging is shown in \cite{FS} to be equivalent to that of F\"{o}llmer-Schweizer (F-S) decomposition of the relevant discounted claim for a particular class of asset price dynamics.

\noi In recent years, there have been phenomenal advancements in application of regime-switching dynamics in finance. The model parameters here are driven by a finite-state continuous time pure jump process whose states represent the stages of business cycles. One can refer to for example DiMasi et al. \cite{DKR}, Guo \cite{Guo}, Elliott et al. \cite{Elliot2005}, and Siu et al. \cite{Siu} for developement of the theory of option pricing in Markov regime switching model. With the locally risk minimizing approach, the European call option price in a regime switching market is shown to satisfy a generalized B-S-M PDE in the work of \cite{DKR} and many others subsequently including \cite{DG} and \cite{BGG}. An extension of the Markov modulated regime switching model in \cite{DKR} appears in  \cite{AG1} and \cite{AG2}, where the holding time of each regime is not restricted to be exponential variable and the regime dynamics follows a semi-Markov process. A semi-Markov process may, in general, exhibit duration dependent transition which a time homogeneous or a time in-homogeneous Markov chain do not capture. An empirical evidence of duration dependent transition of business cycles is presented in \cite{Chang}. In \cite{Bulla} and \cite{DA} the calibrations of a semi-Markov modulated discrete and continuous time models are presented. The sensitivity of the call option price to the calibration error in the transition rate of a semi-Markov process is studied in \cite{GN}. The price of a basket option in a semi-Markov modulated time in-homogeneous market is also found in \cite{TP} to satisfy an extension of the multi-dimensional version of the BSM PDE. In another recent work \cite{BGO}, the option pricing problem in a local volatility model with semi-Markov switching is solved. All the studies mentioned above assume absence of discontinuities in the asset price dynamics. The option price problem with discontinuous asset price setting was initially addressed in \cite{AA, BC, CH}. A Reader may consult \cite{RC} for various aspects of discontinuous asset price dynamics. The studies by Elliott et al. \cite{Elliot2007} and Su et al. \cite{Su} involve jump diffusion models with Markov type regime switching. In Siu \cite{Siu2} the jump intensity is modulated by a continuous-time, finite-state, hidden Markov chain. The approach of F-S decomposition was adopted in \cite{CH, Su} but not in \cite{AA, BC, Elliot2007, Siu2}. Although the differential equation for option price are presented but the arguments for existence of classical solution are absent in the above mentioned works. As per our knowledge, no existing model capture both the features, the discontinuity in the asset price path, and the duration dependent regime switching. 

\noi In this paper we have firstly verified that the asset price model involving a semi-Markov regime-switching and jump discontinuities in paths does not allow arbitrage opportunity. It is well known that the parameters in a model with jump discontinuities should satisfy an additional constraint for ensuring \emph{no arbitrage} (NA). We have obtained such appropriate condition. Furthermore, we have constructed a nontrivial realistic example which satisfies that  condition. Having established the NA, we have considered a general European type path independent contract in which the terminal payoff is a Lipschitz continuous function of the final stock value. This class of payoffs is broad enough to include options like call, put and butterfly. We have taken the approach of F-S decomposition for finding the locally risk minimizing pricing. Although this is a standard approach in the literature, the issues become more subtle when asset prices are allowed to jump. The conditions under which the \emph{pseudo optimal strategy}, obtained from the F-S decomposition, is locally risk minimizing needs a careful checking. Furthermore, the F-S decomposition under the minimal martingale measure may not produce an F-S decomposition under the original measure. We refer to \cite{VV} which nicely highlights this aspect. We have appropriately obtained the F-S decomposition under the market measure to construct the pseudo optimal strategy. Subsequently, the pseudo optimal strategy is shown to be optimal.

\noi From the F-S decomposition we have established that the locally risk minimizing price of a Lipschitz terminal payoff can be expressed as a solution to an appropriate evolution problem. The corresponding equation turns out to be a linear, non-local, degenerate parabolic system of integro-PDEs which, when restricted, gives the BSM PDE. However, this evolution problem does not possess a closed form solution and  is very different from its well-studied special cases. The methodology, used in the literature for settling well-posedness of some of the special cases, is also inadequate to deal with the present price equation. The treatment, adopted in this paper, does also not follow immediately from the existing results in the literature of integro-PDE. In this paper, we establish the existence and uniqueness of the classical solution to the problem using semigroup theory. This is accomplished in two steps. First, the price equation is shown to have a continuous mild solution satisfying an integral equation. Then by studying the integral equation, it is proved that its solution is sufficiently smooth and belongs to the domain of the differential operator in the evolution equation. Thus the mild solution is shown to solve the equation classically. Classical solution is important for creating an optimal hedging. Without estimates on the hedging strategy, which can only be obtained by analyzing the classical solution, one cannot establish the admissibility of the hedging strategy. This is also an essential step for deriving F-S decomposition.

\noi Rest of the paper is organized as follows. Section 2 has four subsections. The asset price dynamics is demonstrated in the first three, whereas the fourth subsection establishes the no arbitrage condition. The option price equation is introduced at the end of Subsection 2.3. There are two subsections in the third section. The locally risk minimizing pricing using  F-S decomposition in an abstract setting is briefly recalled in the first subsection. In Subsection 3.2 we derive the F-S decomposition of a Lipschitz payoff on final stock value under our specific market model. In this part, we have used the existence and uniqueness theorem for classical solution to the option price equation. That theorem is proved to be true in Section 4. Proof of some technical results are added in the appendix. We end this paper with some concluding remarks in Section 5.

\section{Market model}\label{model}
\subsection{Probability space}
\noindent Let $(\Omega, \mathcal{F}, \{\mathcal{F}_t\}_{t\geq0}, P)$ be a complete filtered probability space where the filtration satisfies the usual hypothesis and $\mathcal{X} = \{1, 2, 3,\ldots, k\}$ a finite subset of $\mathbb{R}$.
Let $X_0$, a $\mathcal{X} $ valued random variable, and $Y_0$, a non-negative random variable, be $\mathcal{F}_0$ measurable. We further assume that $\wp$, a Poisson random measure on $[0,\infty)\times [0,\infty)$ with uniform intensity, $W:=\{W_t\}_{_{t\geq0}}$, a standard Wiener process, and $N(dz,dt)$, a Poisson random measure on $\mathbb{R}\times [0,\infty)$ with intensity $\nu(dz)dt$ where $\nu$ is a finite Borel measure, are adapted to the filtration $\{\mathcal{F}_t\}_{t\geq0}$. We take $X_0$, $Y_0$, $\wp$, $W$, and $N(dz,dt)$ as independent.

\noi The above mentioned random variables, processes and measures would be used to model the inherent randomness in the asset price dynamics in a financial market. We first present an SDE for the regimes which is modeled as a semi-Markov process. Since such SDE is not very common in the literature, for the sake of self-containedness, we summarize some related definitions and properties in the following subsection before presenting the asset price dynamics in Subsection 2.3.

\subsection{Semi-Markov process}
We begin this subsection by recollecting the definition of a time-homogeneous semi-Markov process or in short semi-Markov process as we would not consider the  time-inhomogeneous case in this paper.
\begin{defi} \label{SMD} A $\mathcal{X}$-valued process $X=\{X_t\}_{t\ge0}$ defined on $(\Omega, \mathcal{F}, P)$ is called a semi-Markov process if\\
i. for almost every $\omega\in \Omega$, $X$ has a piece-wise constant rcll path with discontinuities at an increasing unbounded positive sequence  $\{T_n(\omega)\}_{n\ge1}$, and\\
ii. $P(X_{T_{n+1}} = j, T_{n+1}- T_{n} \le y \mid (X_{T_0}, T_0),(X_{T_1}, T_1), \ldots ,(X_{T_n}, T_n) )= P(X_{T_{n+1}} = j, T_{n+1}- T_{n} \le y \mid X_{T_n})$ for all $j\in\mathcal{X} $, $y>0$, $n\ge 0$ and for some $T_0 \le 0$, where $X_{T_0}:=X_0$.
\end{defi}

\noi Given a semi-Markov process $X=\{X_t\}_{t\ge 0}$, on $\mathcal{X}$, with transition time sequence $T_0\le 0< T_1<T_2,\ldots$; the instantaneous transition rate function of $X$, if exists, is a collection $\lambda_{ij}: [0, \infty) \to [0, \infty), \forall i \neq j \in \mathcal{X}$, given by
$$\lambda_{ij}(y):=\lim_{\delta \to 0} \frac{1}{\delta} P\left[ X_{T_{n+1}}=j, T_{n+1}-T_n\in (y, y+\delta) |X_{T_n}=i, T_{n+1}-T_n >y \right],$$
and does not depend on $n$. We consider the class of semi-Markov processes which admits the above limit satisfying the following additional conditions.
\begin{itemize}
\item[{\bf A1.}](i) For each $i\neq j \in \mathcal{X}$, $\lambda_{ij}:[0,\infty)\to [0,\infty)$ is a bounded, positive and continuously differentiable function.\\
(ii) If $\la_i(y):=\sum_{j\neq i}\lambda_{ij}(y)$ and $\Lambda_i(y):=\int_0^y\la_i(y) dy$, then $\lim_{y\to \infty}\Lambda_i(y) =\infty$.
\end{itemize}
\noindent Note that the instantaneous transition rates of an irreducible finite-state continuous time Markov chain are only  positive constants. Therefore the assumption (A1) includes the Markov counterpart as a special case. Here we recall the following results from \cite{AG2}.

\begin{prop}\label{prop} Given a collection $\lambda=\{ \lambda_{ij}:[0,\infty)\to (0,\infty) \mid i \neq j \in \mathcal{X}\}$ of bounded measurable maps, satisfying (A1(ii)), there exist a finite interval $\mathbb{I}$ and a piecewise linear maps $h_{\lambda}$ and $g_{\lambda}$ on $\mathcal{X}\times [0,\infty)\times \mathbb{I}$ such that the following hold.
\begin{itemize}
        \item[(i)] The system of coupled stochastic integral equations
        \begin{align}
         X_t &= X_0 + \int_{(0,t]}\int_{\mathbb{I}}h_{\lambda}(X_{u-},Y_{u-},z)\,\wp(dz,du)\label{xdef},\\
         Y_t &= Y_0+t-\int_{(0,t]}\int_{\mathbb{I}}g_{\lambda}(X_{u-},Y_{u-},z)\,\wp(dz,du),\label{ydef}
        \end{align}
\label{Y}
for all $t>0$ has a strong rcll solution $(X,Y)$ for any given $(X_0, Y_0)\in \mathcal{X} \times [0,\infty)$ such that the solution $X$ is a semi-Markov process on $\mathcal{X}$ with $\lambda$ as instantaneous transition rate function and $Y$ gives the age at the current state. More precisely, $Y_t = t-T_{n(t)}$, where $n(t)$ is the number of transitions during $(0,t]$.
\item[(ii)] The infinitesimal generator $\mathcal{A}$ of $(X,Y)$ is given by $\mathcal{A}\varphi (i,y) = \frac{\partial\varphi}{\partial y} (i,y)+ \sum_{j \neq i} \lambda_{ij}(y) \Bigl( \varphi(j,0) - \varphi(i,y)\Bigr)$ for every  function $\varphi: \mathcal{X}\times [0,\infty)\to \mathbb{R}$ with bounded continuous derivatives.
\item[(iii)] Consider $F:[0,\infty)\times \mathcal{X}\rightarrow [0,1]$, defined as $F(y|i):=1-e^{-\Lambda_i(y)}$, where $\Lambda_i$ is as in (A1)(ii). Then under A1(i), $F(\cdot \mid i)$ is a twice differentiable function and is the conditional cumulative distribution function of the holding time of $X$ given that the present state is $i$.
\item[(iv)]  We define $p_{ij}(y):=\frac{\la_{ij}(y)}{\la_i(y)}$ for $j\neq i$ with $p_{ii}(y)=0$ for all $i$ and $y$.  Then for each $i$, $j$ and $y$, $p_{ij}(y)$ denotes the conditional probability of transition to $j$ given that the process transits from $i$ at the age $y$.
\item[(v)] Under A1(i) $f(y|i):=\frac{d}{dy}F(y|i)$ is bounded and continuously differentiable. Moreover, $\la_{ij}(y) = p_{ij}(y)\frac{f(y|i)}{1-F(y|i)}$ holds for all $i\neq j$.
\end{itemize}
\end{prop}
\noi From (i) in the above proposition, $Y_0=-T_0$. However, for the sake of simplicity, we assume that the initial age, i.e., $Y_0$ is zero. This causes no  loss of generality in our treatment. Indeed, since the regimes are assumed to be observed, for valuation of options with maturity $\tau$ one may, in the case of nonzero initial age, set the value of initial time as the initial age $y^*$ (say) instead of zero, so as to make $Y_{y^*}=y^*$, or in other words, $Y_0=0$. Hence, the new maturity time $T$ is $y^*+\tau$. Of course as a consequence $T_0$ (as in Definition \ref{SMD}) is zero and $Y_t\in [0,t]$ for all $t\in [0,T]$. However we do not fix this canonical selection unless it is clearly specified in the subsequent analysis.

\subsection{Asset price dynamics}
In this subsection the mathematical model of a financial market dynamics is presented. We consider a market having two types of securities, one is a risk-less asset, also called money market instrument, and another is a risky asset, called stock. Let $X=\{X_t\}_{t\geq0}$ be a semi-Markov process on the state space $\mathcal{X}$ satisfying A1 and (\ref{xdef})-(\ref{ydef}), $r_t := r(X_t)$ be the spot interest rate of risk-less asset whose price is $B_t$ at time $t$ with $B_0 = 1$. Then we have $B_t = e^{\int_0^tr_udu}$. Now let $S:= \{S_t\}_{t\geq0}$ be the price process of the risky asset which is governed by a semi-Markov modulated jump diffusion model as given below
\begin{equation}
dS_t = S_{t-}\Big(\mu_{t-}dt+\sigma_{t-} dW_t+\int_{\mathbb{R}}\eta(z)N(dz,dt)\Big), \label{E23}
\end{equation}
\noindent where $t>0$, $S_0 > 0$, $\mu_t:=\mu(X_t)$ denotes the drift,  $\sigma_t:=\sigma(t,X_t)$ the volatility coefficient, and a bounded continuous function $\eta:\mathbb{R}\rightarrow (-1,\infty)$ the jump size coefficient. Furthermore, the time-inhomogeneous volatility function $\sigma: [0,\infty)\times \mathcal{X} \to (0, \infty)$ is assumed to be continuous.
\begin{theo}\label{theoSDE}
The SDE \eqref{E23} has an almost sure unique strong solution which is given by
\begin{equation}\label{eq2.4}
S_t = S_0 \exp\left( \int_0^t(\mu(X_{u-})-\frac{1}{2}\sigma(u,X_{u-})^2)du+\int_0^t\sigma(u,X_{u-})dW_u+\int_0^t\int_{\mathbb{R}}\ln(1+\eta(z))N(dz,du)\right).
\end{equation}
Furthermore, the  solution is positive valued and square integrable.
\end{theo}
\noi Proof of the above theorem is deferred to the Appendix.
\begin{coro}\label{coro}
(i) The discounted asset price process $S^*=\{S^*_t\}_{t\ge 0}$ given by $S^*_t:=\frac{S_t}{B_t}=\exp\left(-\int_0^tr_u du\right) S_t $, where $S$ is as in  \eqref{eq2.4}, which is also the solution to \eqref{E23} can be rewritten as $S^*_t=S_0+C_t +G_t$ where $G:= \{G_t\}_{t\geq0}$ is a square integrable martingale with $G_0=0$ and $C:= \{C_t\}_{t\geq0}$ a predictable continuous process of finite variation. (ii) The  conditional quadratic variation process $t\mapsto \langle G\rangle_t$ is strictly increasing on $[0,T]$ almost surely. (iii) $C$ is absolutely continuous w.r.t. $\langle G\rangle$ with a density $\delta_C$ such that the \emph{mean-variance tradeoff process} $\widehat{K}$ given by $\widehat{K}_t:= \int_0^t {\delta_C}^2(t) d \langle G\rangle_t$ has finite expectation i.e., $E\widehat{K}_T <\infty$.
\end{coro}
\proof From \eqref{E23}, we directly get that
\begin{align}\label{s*}
\nonumber dS_t^*=&\exp\left(-\int_0^tr_u du\right) dS_t-S_{t-}\exp\left(-\int_0^tr_u du\right) r_{t-}dt\\
\nonumber    =&(1/B_t) \big(dS_t - S_{t-}r_{t-}dt\big)\\
\nonumber =&(1/B_t) \left(S_{t-}\Big(\mu_{t-}dt + \sigma_{t-}dW_t + \int_{\mathbb{R}}\eta(z)N(dz,dt)\Big) - S_{t-}r_{t-}dt\right)\\
    =&\big(\mu_{t-}-r_{t-}\big)S_{t-}^*dt+\sigma_{t-}S_{t-}^*dW_t+S_{t-}^*\int_{\mathbb{R}}\eta(z)N(dz,dt).
\end{align}
Thus $S^*_t=S_0+C_t +G_t$ where
$$ C_t := \int_0^t \Big(\mu_{u-}-r_{u-}+ \int_{\mathbb{R}}\eta(z) \nu(dz)\Big)S^*_{u-} du, \textrm{ and }~ G_t:= \int_0^t \sigma_{u-} S^*_{u-}dW_u+ \int_0^t S^*_{u-} \int_{\mathbb{R}}\eta(z)\tilde{N}(dz,du),
$$
where $S$ is as in \eqref{eq2.4} and  $\tilde{N}(dz,du) := N(dz,du)-\nu(dz)du$ is the compensated Poisson random measure. The square integrability of $G$ follows from the square integrability of $S$. Thus the conditional quadratic variation of $G$ is given by
$$ \langle G\rangle_t= \int_0^t \sigma_{u-}^2 {S^*_{u-}}^2du+ \int_0^t {S^*_{u-}}^2 \int_{\mathbb{R}}\eta(z)^2\nu(dz)du.
$$
Since $\sigma$ is assumed to be strictly positive, $\langle G\rangle$ is strictly increasing. The density $\delta_C$ is defined by
$$\delta_C(t)=\frac{dC_t}{d\langle G\rangle_t}= \frac{\mu_{t-}-r_{t-}+ \int_{\mathbb{R}}\eta(z) \nu(dz)}{S^*_{t-}\left(\sigma_{t-}^2 + \int_{\mathbb{R}}\eta(z)^2\nu(dz)\right)}.$$
Thus
\begin{align*}
E \widehat{K}_T &= E \int_0^T \frac{\left(\mu_{t-} -r_{t-}+ \int_{\mathbb{R}}\eta(z) \nu(dz)\right)^2}{\sigma_{t-}^2 + \int_{\mathbb{R}}\eta(z)^2\nu(dz)}dt <\infty
\end{align*}
since $\eta$ is bounded, $\nu$ is finite and $\inf_{[0,T]\times \mathcal{X} }\sigma(t,i) >0$ for finite $T$.
\qed

\noi The results obtained in the above corollary is stronger than the so called \emph{Structure Condition} (SC) \cite{Sch2001}. We would use these results in the next section.

\noi We would end this subsection with the following observations. The Dynkin's formula states that if $\{A_t\}_{t\ge 0}$ is the infinitesimal generator of $\{(S_t,X_t,Y_t)\}_{t\ge 0}$, then $t\mapsto \psi( S_t,X_t,Y_t)-\psi(S_0,X_0,Y_0)-\int_0^t A_u\psi(S_{u-},X_{u-},Y_{u-})du$
is an $\{\mathcal{F}_t\}_{t\geq 0}$-martingale for any $\psi\in C_c^\infty$. By denoting the above martingale by $\{M_t\}_{t\geq0}$, we get
\begin{equation*}
    \psi(S_t,X_t,Y_t)=\psi(S_0,X_0,Y_0)+\int_0^t A_u\psi(S_{u-},X_{u-},Y_{u-})du + M_t.
\end{equation*}
\noindent From (\ref{xdef}), (\ref{ydef}) and \eqref{E23}  and using Proposition \ref{prop}(ii) we get
\begin{align*}
    A_t\psi(s,i,y)= &\Big(\mu(i)s\frac{\partial}{\partial s}+\frac{1}{2}\sigma^2(t,i)s^2\frac{\partial^2}{\partial s^2}+\frac{\partial}{\partial y}\Big)\psi(s,i,y) + \sum_{j\neq i}\lambda_{ij}(y)\big(\psi(s,j,0)-\psi(s,i,y)\big)\\
    &+ \int_{\mathbb{R}}\big(\psi(s(1+\eta(z)),i,y)-\psi(s,i,y)\big)\nu(dz).
\end{align*}
In Section 3 we consider, with an appropriate terminal condition at time $T$, the following integro-partial differential equations (IPDE) for $(t,s,i,y)\in \D:=(0,T)\times (0,\infty)\times \X \times (0,\infty)$
\begin{align}
    \nonumber & \Big(\frac{\partial}{\partial t}+A_t +\big(r(i)-\mu(i)+\beta_1(t,i)\big)s\frac{\partial}{\partial s}\Big)\varphi(t,s,i,y)\\
    &+  \int_{\mathbb{R}} \big(\beta_2(t,z,i)-1\big) \big(\varphi(t, s(1+\eta(z)), i, y)-\varphi(t, s, i, y)\big)\nu(dz) = r(i)\varphi(t, s, i, y) \label{L24}
\end{align}
\noindent  where the continuous functions $\beta_1(t,i)$ and $\beta_2(t,z,i)$ do not depend on $\varphi$ and are yet to be chosen.
	
	\subsection{No arbitrage}
	
An arbitrage is an indication of the instability in the market. A market is said to have an arbitrage when it enables an investor to get positive profit without an initial capital and possibility of loss.  So we need to check whether this model has no arbitrage (NA) under a reasonably large class of admissible strategies. From Theorem VII.2c.2 of \cite{Shir}, one obtains that the existence of an equivalent martingale measure (EMM) implies NA in the sense of NFLVR (no free lunch with vanishing risk) under a class of admissible strategies. Thus to ensure an arbitrage free model, exhibition of an EMM is called for.
An EMM is commonly constructed using a Radon-Nikodym process involving a Dol\'{e}ans-Dade exponential. However, in general, a Dol\'{e}ans-Dade exponential of a martingale is a super martingale giving rise to a sub-probability measure. A Novikov's type condition, if satisfied, resolves this deficit by asserting that the stochastic exponential is a martingale. We appeal to the results in \cite{PR} for a similar condition in the jump diffusion setting in order to show that a sufficiently large class of  Dol\'{e}ans-Dade exponential are martingales in the following lemma.

\begin{lem} \label{lem 1}
Let $Z = \{Z_t\}_{t\in[0,T]}$ be an adapted process, defined as
\begin{equation}
Z_t := \exp\left(\int_0^t\phi_udW_u-\frac{1}{2}\int_0^t\phi_u^2du+\int_0^t\int_{\mathbb{R}}\ln \Gamma_u(z)N(dz,du) - \int_0^t\int_{\mathbb{R}}\Big(\Gamma_u(z)-1\Big)\nu(dz)du\right), \label{L25}
\end{equation}
\noindent where $\phi = \{\phi_t\}_{t\in [0,T]}$ and $\Gamma = \{\Gamma_t(\cdot)\}_{t\in[0,T]}$ are predictable and bounded processes with $\Gamma > 0$. Then $Z$ is a positive martingale under $P$ with $Z_0 = 1$.
\end{lem}

\proof From \eqref{L25}, it is obvious that $Z > 0$ with $Z_0 = 1$. We derive the following
\begin{align*}
    \Delta Z_t =& Z_t - Z_{t-}\\
    =&\exp\left(\int_0^t\phi_udW_u - \frac{1}{2}\int_0^t\phi_u^2du + \int_0^t\int_{\mathbb{R}}\big(\Gamma_u(z)-1\big)\nu(dz)du\right) \exp\left(\int_{[0,t]}\int_{\mathbb{R}}\ln \Gamma_u(z)N(dz,du)\right)\\
    &-\exp\left(\int_0^t\phi_udW_u -\frac{1}{2}\int_0^t\phi_u^2du + \int_0^t\int_{\mathbb{R}}\big(\Gamma_u(z)-1\big)\nu(dz)du\right) \exp\left(\int_{[0,t)}\int_{\mathbb{R}}\ln \Gamma_u(z)N(dz,du)\right)\\
    =&Z_{t-}\left(\int_{\mathbb{R}}\big(\Gamma_t(z)-1\big)N(dz,\{t\})\right).
\end{align*}
We define $y_t:=\int_0^t\phi_udW_u-\frac{1}{2}\int_0^t\phi_u^2du+\int_0^t\int_{\mathbb{R}}\ln \Gamma_u(z)N(dz,du) - \int_0^t\int_{\mathbb{R}}\big(\Gamma_u(z)-1\big)\nu(dz)du$. Hence, $\Delta y_t= \int_{\mathbb{R}}\ln \Gamma_t(z)N(dz,\{t\})$ and an application of It\^{o} formula on $Z_t=\exp(y_t)$ gives,
\begin{align}
    \nonumber Z_t-Z_0=&\int_0^tZ_{u-}dy_u + \frac{1}{2}\int_0^tZ_{u-}d[y]_u^c + \sum_{0<u\leq t}\Big(Z_u-Z_{u-}-Z_{u-}  \Delta y_u\Big)\\
    \nonumber
    Z_t-1= &\int_0^tZ_{u-}\phi_udW_u-\int_0^tZ_{u-}\frac{1}{2}\phi_u^2du+\int_0^tZ_{u-}\int_{\mathbb{R}}\ln \Gamma_u(z)N(dz,du) -\int_0^tZ_{u-}\int_{\mathbb{R}}\big(\Gamma_u(z)-1\big)\nu(dz)du\\ \nonumber
    & + \frac{1}{2}\int_0^tZ_{u-}\phi_u^2du+\sum_{0<u\leq t} Z_{u-} \int_\mathbb{R}\Big(\Gamma_u(z)-1-\ln \Gamma_u(z)\Big)N(dz,\{u\})\\
    Z_t=&1+\int_0^t Z_{u-}\phi_udW_u+\int_0^tZ_{u-}\int_{\mathbb{R}}\big(\Gamma_u(z)-1\big)\tilde{N}(dz,du).\label{L26}
\end{align}
\noindent From above, we can see that $Z$ satisfies
\begin{equation}\label{zz}
dZ_t=Z_{t-}d\hat{M}_t, ~~~  Z_0 =1
\end{equation}
\noindent where $\hat{M}:=\{\hat{M}_t\}_{t\ge 0}$ is a $P-$local martingale given by
$$\hat{M}_t:= \int_0^t \phi_udW_u + \int_0^t \int_{\mathbb{R}} \big(\Gamma_u(z)-1\big) \tilde{N}(dz,du).$$
However, our assumptions on $\phi$ and $\Gamma$ imply that $\hat{M}$ is a  square integrable martingale with jump size ($\Delta M$) greater than $-1$.
Thus in view of \eqref{zz},  $Z$ is the Dol\'{e}ans-Dade exponential of $M$ and satisfies all the conditions given in Theorem 9 of \cite{PR}. Hence, applying that theorem we conclude that $Z$ is a positive martingale. \qed

\noindent The following lemma, which is essentially borrowed from the Theorem 3.2 of \cite{CH}, presents the change in law of underlying processes under a new measure $Q$, constructed via the Radon-Nikodym process $Z$ as above. The result can be viewed as a consequence of a version of the Girsanov theorem. This lemma is useful to pin down an EMM for our specific model.
\begin{lem}
\label{lem 2}
Let $Q$ be defined on $\mathcal{F}_T$ by $\frac{dQ}{dP}=Z_T$, where $Z$ is as in \eqref{L25}. Then the process $\tilde{W} := W - \int_0^\cdot \phi_u du$ is a Wiener process under $Q$ and
\begin{equation*}
\int_0^\cdot\int_{\mathbb{R}}\Big(\Gamma_u(z)-1\Big)\Big(N(dz,du)-\Gamma_u(z)\nu(dz)du\Big)
\end{equation*}
\noindent is a $Q-$martingale with respect to its natural filtration. The compensator measure of $N(dz,dt)$ under $Q$ is given by $\tilde{\nu}(dz,dt) := \Gamma_t(z)\nu(dz)dt$.
\end{lem}
\noindent Lemma \ref{lem 2} implies that $\tilde{M}(dz,dt) := N(dz,dt) - \tilde{\nu}(dz,dt)$ is a compensated Poisson random measure with respect to the measure $Q$. We rewrite the SDE \eqref{s*} using $\tilde{W}$, $\tilde{M}$ and the unspecified processes $\phi$ and $\Gamma$ to get
\begin{align}
    \nonumber dS_t^*=& \big(\mu_{t-}-r_{t-}\big)S_{t-}^*dt + \sigma_{t-}S_{t-}^*\big(d\tilde{W}_t+\phi_{t}dt\big)+S_{t-}^*\int_{\mathbb{R}}\eta(z)\big(\tilde{M}(dz,dt)+\tilde{\nu}(dz,dt)\big)\\
    =& \left(\mu_{t-}-r_{t-}+\sigma_{t-}\phi_t+\int_{\mathbb{R}}\eta(z)\Gamma_t(z)\nu(dz)\right)S_{t-}^*dt+\sigma_{t-} S_{t-}^* d\tilde{W}_t + S_{t-}^* \int_{\mathbb{R}}\eta(z)\tilde{M}(dz,dt). \label{L27}
\end{align}
\noindent Now we wish to specify $\Gamma$ and $\phi$ so as to make $S^*$ a martingale under $Q$. In view of Lemma \ref{lem 2}, it is possible only when the drift term in \eqref{L27} is zero. Thus we have
\begin{equation}
\mu_{t-}-r_{t-}+\sigma_{t-}\phi_t+\int_{\mathbb{R}}\eta(z)\Gamma_t(z)\nu(dz) = 0. \label{L28}
\end{equation}
This is a single equation with two unknowns. Hence \eqref{L28} leads to many different possibilities
corresponding to different pairs of $(\phi,\Gamma)$ satisfying \eqref{L28}. We would like to select one which satisfies an additional relation such that \eqref{L26} can be
represented as
\begin{equation}
    dZ_t= J_{t-}Z_{t-}(\sigma_{t-}dW_t + \int_{\mathbb{R}}\eta(z)\tilde{N}(dz,dt)), Z_0=1 \label{L29}
\end{equation}
\noindent for some bounded adapted process $J:=\{J_t\}_{t\ge 0}$. Now by comparing \eqref{L26} and \eqref{L29}, we get
\begin{equation*}
    \phi_t =  J_{t-}\sigma_{t-}, ~~~~\textrm{and ~} \Gamma_t(z) - 1=  J_{t-}\eta(z) .
\end{equation*}
\noindent Hence by substituting above in \eqref{L28}, we obtain
\begin{align*}
     J_t\sigma_t^2 =& \ r_t - \mu_t - \int_{\mathbb{R}}\eta(z)(1+ J_t\eta(z))\nu(dz)\\
    =& \ r_t - \mu_t - \int_{\mathbb{R}}\eta(z)\nu(dz) -  J_t\int_{\mathbb{R}}\eta^2(z)\nu(dz)
\end{align*}
\noindent for all $t$. Therefore $ J_t$ can be written as
\begin{equation*}
     J_t = \frac{r_t - \mu_t - \int_{\mathbb{R}}\eta(z)\nu(dz)}{\sigma^2_t+\int_{\mathbb{R}}\eta^2(z)\nu(dz)},
\end{equation*}
\noindent which results in
\begin{align}
    \begin{rcases}
    \Gamma_t(z) &= \frac{r_{t-} - \mu_{t-} - \int_{\mathbb{R}}\eta(z)\nu(dz)}{\sigma^2_{t-}+\int_{\mathbb{R}}\eta^2(z)\nu(dz)}\eta(z) +1\\
    \phi_t &= \frac{r_{t-} - \mu_{t-} - \int_{\mathbb{R}}\eta(z)\nu(dz)}{\sigma^2_{t-}+\int_{\mathbb{R}}\eta^2(z)\nu(dz)}\sigma_{t-}. \label{L30}
    \end{rcases}
\end{align}
\noindent It is easy to see that under the assumption
\begin{itemize}
\item[\textbf{A2.}]
$ \frac{r(i) - \mu(i) - \int_{\mathbb{R}}\eta(z)\nu(dz)}{\sigma^2(t,i)+\int_{\mathbb{R}}\eta^2(z)\nu(dz)}\eta(z) > -1$
for all $t\in[0,T]$, $i\in\mathcal{X}, z\in\mathbb{R}$,
\end{itemize}
the conditions on $\phi$ and $\Gamma$ in Lemma \ref{lem 1} hold true. We assume (A2) throughout the paper and illustrate some of its implications in Remark \ref{rem1}. Thus under (A2), using \eqref{L30} and Lemma \ref{lem 1}, the measure $Q$ as in Lemma \ref{lem 2} is a probability measure equivalent to $P$. Furthermore, the substitution of \eqref{L30} in \eqref{L27}, gives
\begin{equation*}
    dS^*_t= S^*_t\Big(\sigma(t,X_{t-})d\tilde{W}_u + \int_{\mathbb{R}}\eta(z)\tilde{M}(dz,dt)\Big),
\end{equation*}
a Dol\'{e}ans-Dade exponential, which is, by the virtue of Lemma \ref{lem 2}, and Theorem 9 of \cite{PR}, a positive martingale under $Q$. Thus we have proved the following theorem.
\begin{theo}
\label{theoq}  Under $(A2)$, the substitution of the values of $\phi_t$ and $\Gamma_t(z)$ from \eqref{L30} in \eqref{L25} leads to an equivalent martingale measure $Q$ which is given in Lemma \ref{lem 2}. Hence the market is arbitrage free.
\end{theo}
\begin{rem} \label{rem1} An assumption similar to (A2) is standard in the literature. We refer to \cite{Siu} for an instance. But in the BSM model, or in its regime switching generalizations no such assumption is required. It is evident that the BSM model is a special case of the present model in which the asset price is a continuous function of time, or in other words, $\eta$ is identically zero. It is important to note that if $\eta \equiv 0$, (A2) holds for any choice of $r$, $\mu$ and $\sigma$ and thus imposes no further constraint on any model parameter. However, for a nontrivial $\eta$, (A2) puts bounds on the drift coefficient $\mu$. That is,
$
\left(\mu(i)-\left(r(i) - \int_{\mathbb{R}}\eta(z)\nu(dz) \right)\right)\eta(z) < \sigma^2(t,i)+\int_{\mathbb{R}}\eta^2(z)\nu(dz)$ or
\begin{align*}
&\left(r(i) - \int_{\mathbb{R}}\eta(z)\nu(dz) \right) - \left(\sigma^2(t,i)+\int_{\mathbb{R}}\eta^2(z)\nu(dz)\right)/\max(0,-\eta(z))\\
&< \mu(i)< \left(r(i) - \int_{\mathbb{R}}\eta(z)\nu(dz) \right) + \left(\sigma^2(t,i)+\int_{\mathbb{R}}\eta^2(z)\nu(dz)\right)/\max(0,\eta(z))
\end{align*}
for all $i,t$, and $z$. In view of the fact that any real risky asset has greater drift value than the ideal bank rate, it is important to cross check if the above mentioned upper bound leads to an unrealistic model assumption. In order to illustrate the implication of this upper bound, we consider a nontrivial example where $\eta(z)= \max(\min(z,1),-\frac{1}{2})$, $\nu$ is the Lebesgue measure on $[-\frac{1}{2},1]$. Then $\int_{\mathbb{R}}\eta(z)\nu(dz)$=$\int_{\mathbb{R}}\eta^2(z)\nu(dz)$=3/8. Therefore, (A2) is true if $r(i) - 3/8  - 2(\sigma^2(t,i)+3/8 ) <\mu(i) < r(i) - 3/8  + \sigma^2(t,i)+3/8$ for all $i$, and $t$. This holds if $r(i)<9/8$ and $-2 \min_{[0,T]}\sigma^2(t,i)\le \mu(i) <r(i) +\min_{[0,T]}\sigma^2(t,i)$ for all $i$. These bounds are clearly not unrealistic.
\end{rem}

\section{Pricing and Optimal Hedging}\label{section for pricing}

\subsection{Locally Risk Minimizing Approach}
\noindent Let $\xi_t$ and $\varepsilon_t$ be the number of units invested in assets with prices $S_t$ and $B_t$ respectively at time $t$. The value of the resulting portfolio at time $t$ is given by
\begin{equation*}
    V_t := \xi_tS_t + \varepsilon_tB_t.
\end{equation*}
\noindent An admissible strategy is defined to be a predictable process $\pi = \{\pi_t = (\xi_t,\varepsilon_t)\}_{t\in [0, T]}$ which satisfies the following conditions

\noindent (i) $\int_0^T\xi_t^2d\langle G\rangle_t + E(\int_0^T |\xi_t| |dC_t|)^2< \infty$, where $G$ and $C$ are as in Corollary \ref{coro}.

\noindent (ii) $E(\varepsilon_t^2) < \infty,$

\noindent (iii) $\exists a > 0$ s.t. $P(V_t \geq -a, \forall t\in [0,T]) = 1.$
\noindent It is shown in \cite{FS} that if the market is arbitrage free, under some conditions the existence of an \emph{optimal hedging} for replicating an $\mathcal{F}_T$ measurable finite-variance payoff $H$, is equivalent to the existence of F-S decomposition of the discounted payoff $H^* := B_T^{-1}H$ in the form
\begin{equation*}
    H^* = H_0 + \int_0^T\xi_t^{H^*}dS_t^* + L_T^{H^*}
\end{equation*}
\noindent where $H_0 \in L^2(\Omega,\mathcal{F}_0,P)$, $L_{H^*}: = \{L_t^{H^*}\}_{t\in [0,T]}$ is a square integrable martingale starting with zero and orthogonal to the martingale part of $S$, and $\xi^{H^*} = \{\xi_t^{H^*}\}_{t\in [0,T]}$ satisfies (i).
A set of sufficient conditions (See Theorem 3.3 of \cite{Sch2001} for more details), indicated in the above sentence are given below.\\
(i) The conditional quadratic variation of the martingale part of $S^*$ , i.e.,  $\langle G\rangle$ is  strictly increasing,\\
(ii) $t\mapsto C_t$ is continuous, and\\
(iii) $E\widehat{K}_T <\infty$.\\
These conditions do hold in our setting. Indeed each of these conditions is established in Corollary \ref{coro}. In \cite{FS}, it is further asserted that the \emph{optimal hedging} $\pi = (\xi, \varepsilon)$ is given by
\begin{align*}
    \xi_t &:= \xi^{H^*}_t,\\
    V_t^* &:= H_0 + \int_0^t\xi_udS_u^* + L_t^{H^*},\\
    \varepsilon_t &:= V_t^* - \xi_tS_t^*,
\end{align*}
\noindent and $V_t:=B_tV_t^*$ represents the locally risk minimizing price at time $t$ of the contract having a terminal payoff $H$. Hence for the proposed market model, F-S decomposition is the key thing to settle the pricing and hedging problems under the locally risk minimizing approach. In this connection we recall that in the earlier section we have constructed an equivalent martingale measure (EMM) in order to prove that the proposed model does not admit arbitrage in the sense of NFLVR. Once such a risk neutral measure is obtained, one can of course obtain a \emph{no arbitrage} price of a contract by taking conditional expectation of the discounted contingent claim w.r.t. the EMM. It is also not difficult to show that $Q$, the EMM obtained in the earlier section is indeed the \emph{minimal martingale measure} (MMM).
Thus the price obtained thereby is the locally risk minimizing price. Often it is also tempting to apply the It\={o}'s rule on the price function to derive a differential equation for the price function. However, apparently it is not obvious that this function has sufficient regularity required for an application of It\={o}'s rule. Besides, those derivations do not give an answer to the hedging problem also. For these reasons we have avoided that path in this paper. Instead we have first considered an ad-hoc equation and have established the existence and uniqueness of classical solution in Section 4. The analysis of the equation is deferred only to retain the focus on option pricing in the present section. In the next subsection we use that classical solution to derive the desired F-S decomposition. Thereby we conclude that the equation under consideration is indeed the price equation. In this way we settle both the pricing and hedging problem.

\subsection{Derivation of F-S decomposition}

\noindent In order to price an option with a terminal payoff $H = K(S_T)$ where $K:[0,\infty)\to \mathbb{R}$ is a Lipschitz continuous function, we derive the F-S decomposition directly under $P$. To this end we consider the evolution problem given by \eqref{L24} with the terminal condition
\begin{equation}\label{term}
\varphi(T, s, i, y) = K(s).
\end{equation}
It is important to note that the functional parameters $\beta_1(t,i)$ and $\beta_2(t,z,i)$ are yet to be chosen. For the time being, we assume that there is a nonempty collection of pairs of $(\beta_1, \beta_2)$ so that the above evolution problem \eqref{L24}, \eqref{term} has a unique classical solution with at most linear growth for each pair in that collection. We denote a solution by $\varphi \in C^{1,2,1}(\D)$. The above assertion is precisely stated in Theorem \ref{theo45} and is proved for a particular relevant choice of $(\beta_1, \beta_2)$ in the next section.

\noindent For the sake of brevity of notations we denote $\varphi(t, S_t, X_t, Y_t)$ and $\frac{\partial\varphi}{\partial s}(t, S_{t-}, X_{t-}, Y_{t-})$ by  $\varphi_t$ and  $\frac{\partial\varphi_{t-}}{\partial s}$ respectively. In view of the F-S decomposition mentioned in the previous subsection, we are looking for a  predictable process $\xi$, such that
\begin{align}\label{Lt}
L:= \int_0^\cdot \big[d(\frac{\varphi_u}{B_u}) - \xi_udS_u^*\big]
\end{align}
becomes square integrable $P-$martingale and orthogonal to $\bar{M}$, the martingale part of $S^*$. It is easy to see that $\bar{M}_t = \int_0^t\sigma_{u-}S_{u-}^*dW_u + \int_0^tS_{u-}^*\int_\mathbb{R}\eta(z)\tilde{N}(dz, du)$ for all $t\ge 0$.
If such process $\xi$  is found, in view of Subsection 3.1, that constitutes the optimal hedging for $\varphi_T$ i.e., $K(S_T)$ and $\varphi_t$ gives the  locally risk minimizing price at time $t$ of a European style contract with the final payoff $K(S_T)$ at time $T$.
However, there may not exist any such $\xi$ corresponding to an arbitrary choice of $\beta_1$ and $\beta_2$. Hence, in the due course we would also pin down an appropriate choice of $\beta_1$ and $\beta_2$ which allow existence of such $\xi$. By applying It\^{o} formula on $\varphi(t, S_t, X_t, Y_t)$, and using \cref{L24,s*}  we get
\begin{align}
    \nonumber dL_t =& d(\frac{\varphi_t}{B_t}) - \xi_tdS_t^*\\
    \nonumber =& \Big(\frac{-r_{t-}}{B_t}\Big)\varphi_{t-} dt +  \frac{1}{B_t}\Big(\frac{\partial \varphi_{t-}}{\partial t}+A_{t-}\varphi_{t-}\Big)dt +  \frac{1}{B_t}\Big(\frac{\partial\varphi_{t-}}{\partial s}\sigma_{t-} S_{t-} dW_t\Big) \\
    \nonumber & \tab + \frac{1}{B_t}\int \Big(\varphi(t, S_{t-}(1+\eta(z)), X_{t-}, Y_{t-}) -\varphi_{t-}\Big)\tilde{N}(dz,dt) + \frac{1}{B_t}d\widehat{M}_t\\
    \nonumber & \tab \tab - \xi_t\Big(\big(\mu_{t-} - r_{t-}\big)S^*_{t-}dt + \sigma_{t-}S^*_{t-}dW_t + S^*_{t-}\int \eta(z) N(dz,dt)\Big)\\
    \nonumber =&  \Big[\big(\mu_{t-}-r_{t-}-\beta_1(t,X_{t-})\big)S_{t-}^*\frac{\partial\varphi_{t-}}{\partial s}- \int_{\mathbb{R}}\Big(\frac{\beta_2(t,z,X_{t-}) - 1}{B_t}\Big)\Big(\varphi(t, S_{t-}(1+\eta(z)), X_{t-}, Y_{t-})-\varphi_{t-}\Big)\nu(dz) \\
    \nonumber
    &  - \xi_t\big(\mu_{t-}-r_{t-})S_{t-}^*  - \xi_tS_{t-}^*\int_{\mathbb{R}}\eta(z)\nu(dz)\Big]dt + \big(\sigma_{t-}S_{t-}^* \frac{\partial\varphi_{t-}}{\partial s} - \xi_t\sigma_{t-} S_{t-}^*\big)dW_t\\
    &  + \int_{\mathbb{R}}\Big(\frac{\varphi(t, S_{t-}(1+\eta(z)), X_{t-}, Y_{t-})-\varphi_{t-}} {B_t}-\xi_tS_{t-}^*\eta(z)\Big)\tilde{N}(dz,dt) + \frac{1}{B_t}d\widehat{M}_t \label{e}
\end{align}
\noindent where $\widehat{M}_t :=\int_0^t \int_{\mathbb{R}}\big(\varphi(u, S_{u-}, X_{u-} + h(X_{u-}, Y_{u-}, z), Y_{u-} - g(X_{u-}, Y_{u-}, z))-\varphi_{u-}\big)\hat{\wp}(dz,du)$ and  $\hat{\wp}(dz,dt):=\wp(dz,dt)-dtdz$, the compensated Poisson random measure of $\wp$. Hence the local martingale part of $L$ is equal to
\begin{align*}
    &\int_0^\cdot\Big[\sigma_{u-}S_{u-}^*\Big(\frac{\partial\varphi_{u-}}{\partial s} - \xi_u\Big)dW_u\\
    &\hspace*{1 in}+ \int_{\mathbb{R}}\Big(\frac{\varphi(u, S_{u-}(1+\eta(z)), X_{u-}, Y_{u-})-\varphi_{u-}}{B_u}-\xi_uS_{u-}^*\eta(z)\Big)\tilde{N}(dz,du)
    + \frac{1}{B_u}d\widehat{M}_u\Big].
\end{align*}

\noindent Now we find a suitable $\xi$ such that $L$ becomes orthogonal to $\bar{M}$, i.e., $\langle L,\bar{M}\rangle_t = 0$ for all positive $t$, where $\langle , \rangle$ denotes the conditional quadratic covariation. We note that the quadratic covariation
\begin{align*}
    d[L,\bar{M}]_t =& \sigma_{t-}^2S_{t-}^{*2}\big(\frac{\partial\varphi_{t-}}{\partial s} -\xi_t\big)dt\\
    &+ S_{t-}^*\int_{\mathbb{R}}\Big(\frac{\varphi(t, S_{t-}(1+\eta(z)), X_{t-}, Y_{t-})-\varphi_{t-}} {B_t}\eta(z)-\xi_tS_{t-}^*\eta^2(z)\Big)N(dz,dt).
\end{align*}
\noindent Therefore, the conditional quadratic covariation
\begin{align*}
    d\langle L,\bar{M}\rangle_t =& \sigma_{t-}^2S_{t-}^{*2}\big(\frac{\partial\varphi_{t-}}{\partial s} -\xi_t\big)dt\\
    &+ S_{t-}^*\int_{\mathbb{R}}\Big(\frac{\varphi(t, S_{t-}(1+\eta(z)), X_{t-}, Y_{t-})-\varphi_{t-}} {B_t}\eta(z)-\xi_tS_{t-}^*\eta^2(z)\Big)\nu(dz)dt.
\end{align*}
\noindent Thus $\langle L,\bar{M}\rangle = 0$ if for all positive $t$
\begin{align*}
    \sigma_{t-}^2S_{t-}^{*2}\frac{\partial\varphi_{t-}}{\partial s} + \frac{S_{t-}^*}{B_t}\int_{\mathbb{R}} \Big(\varphi(t, S_{t-}(1+\eta(z)), X_{t-}, Y_{t-})-\varphi_{t-}\Big)\eta(z) \nu(dz)&\\
    = \sigma_{t-}^2S_{t-}^{*2} \xi_t + S_{t-}^{*2}\xi_t\int_{\mathbb{R}}\eta^2(z)\nu(dz)&
\end{align*}
\noindent holds. Therefore, for all $t > 0$, $\langle L,\bar{M}\rangle_t$ is zero if $\xi_t$ is chosen as
\begin{equation}\label{f}
  \xi_t=  \frac{\sigma_{t-}^2\frac{\partial\varphi_{t-}}{\partial s} + \frac{1}{S_{t-}} \int_{\mathbb{R}} \Big(\varphi(t, S_{t-}(1+\eta(z)), X_{t-}, Y_{t-})-\varphi_{t-}\Big)\eta(z) \nu(dz)}{ \left(\sigma_{t-}^2  + \int_{\mathbb{R}}\eta^2(z)\nu(dz)\right)}.
\end{equation}

\noindent Thus we have essentially proved that the above choice of $\xi$ makes the local martingale part of $L$ orthogonal to $\bar{M}$, irrespective of the choice of $\beta_1(t,i)$ and $\beta_2(t,z,i)$. However, we have not yet established existence of a particular pair $(\beta_1, \beta_2)$ for which $L$ is a square integrable martingale. It is evident that for ensuring $L$ to be a local martingale, the coefficient of $dt$ term in \eqref{e} should be zero, that is,
\begin{align*}
  0=  & \big(\mu_{t-}-r_{t-}-\beta_1(t,X_{t-})\big)\frac{\partial\varphi_{t-}}{\partial s} - \big(\mu_{t-}-r_{t-}\big)\xi_t \\
   & - \int_{\mathbb{R}}\frac{\beta_2(t,z,X_{t-}) - 1}{S_{t-}}
    \Big(\varphi(t, S_{t-}(1+\eta(z)), X_{t-}, Y_{t-})-\varphi_{t-}\Big)\nu(dz) - \xi_t\int_{\mathbb{R}}\eta(z)\nu(dz).
\end{align*}
\noindent This follows if we have
\begin{align}
    \nonumber \left( (\mu_{t-}-r_{t-}) + \int_{\mathbb{R}}\eta(z)\nu(dz)\right)\xi_t  =& \big(\mu_{t-}-r_{t-}-\beta_1(t,X_{t-})\big)\frac{\partial\varphi_{t-}}{\partial s} -  \int_{\mathbb{R}} \frac{\beta_2(t,z,X_{t-}) - 1}{S_{t-}} \times \\
   &   \Big(\varphi(t, S_{t-}(1+\eta(z)), X_{t-}, Y_{t-})-\varphi_{t-}\Big)\nu(dz). \label{g}
\end{align}

\noindent Using the expression of $\xi_t$ as in \eqref{f}, the above can be rewritten as
\begin{align*}
    &\Big(\mu_{t-}-r_{t-}+\int_{\mathbb{R}}\eta(z)\nu(dz)\Big)\sigma_{t-}^2 \frac{\partial\varphi_{t-}}{\partial s}\\
    &+ \ \Big(\mu_{t-}-r_{t-}+\int_{\mathbb{R}}\eta(z)\nu(dz)\Big)\frac{1}{S_{t-}} \int_{\mathbb{R}}\Big(\varphi(t, S_{t-}(1+\eta(z)), X_{t-}, Y_{t-})-\varphi_{t-}\Big)\eta(z)\nu(dz)\\
    =& \ \Big(\sigma_{t-}^2+\int_{\mathbb{R}}\eta^2(z)\nu(dz)\Big)\Big(\mu_{t-}-r_{t-}-\beta_1(t,X_{t-})\Big)\frac{\partial\varphi_{t-}}{\partial s} \\
    &  - \Big(\sigma_{t-}^2+\int_{\mathbb{R}}\eta^2(z)\nu(dz)\Big)\frac{1}{S_{t-}} \int_{\mathbb{R}} \Big(\beta_2(t,z,X_{t-})-1\Big)\Big(\varphi(t, S_{t-}(1+\eta(z)), X_{t-}, Y_{t-}) -\varphi_{t-}\Big)\nu(dz).
\end{align*}

\noindent We rearrange the terms to get
\begin{align*}
    &\left(\Big(\mu_{t-}-r_{t-}+\int_{\mathbb{R}}\eta(z)\nu(dz)\Big)\sigma_{t-}^2-(\mu_{t-}-r_{t-})\int_{\mathbb{R}}\eta^2(z)\nu(dz)\right.\\
    & \left.+ \ \beta_1(t,X_{t-})\Big(\sigma_{t-}^2+\int_{\mathbb{R}}\eta^2(z)\nu(dz)\Big) - (\mu_{t-}-r_{t-})\sigma_{t-}^2\right)\frac{\partial\varphi_{t-}}{\partial s}\\
    &=-\frac{1}{S_{t-}}\int_{\mathbb{R}}\left(\Big(\mu_{t-}-r_{t-}+\int_{\mathbb{R}}\eta(z)\nu(dz)\Big)\eta(z) + \Big(\sigma_{t-}^2+\int_{\mathbb{R}}\eta^2(z)\nu(dz)\Big) \left(\beta_2(t,z,X_{t-})-1\right) \right)\times \\
    & \ \ \ \ \ \ \ \ \ \ \ \ \ \ \ \ \  \Big(\varphi(t, S_{t-}(1+\eta(z)), X_{t-}, Y_{t-}) -\varphi_{t-}\Big)\nu(dz).
\end{align*}
\noindent The above identity holds true irrespective of the solution function $\varphi$ if $\beta_1$ and $\beta_2$ are such that the coefficients on both sides vanish for all time $t$. A direct calculation shows that such $\beta_1$ and $\beta_2$ exist and are given by
\begin{align}\label{beta}
    \begin{rcases}
    \beta_1(t,i) = \ \frac{\big(\mu(i)-r(i)\big)\int_{\mathbb{R}}\eta^2d\nu - \sigma(t,i)^2\int_{\mathbb{R}}\eta d\nu}{\sigma^2(t,i) + \int_{\mathbb{R}}\eta^2 d\nu}\\
    \beta_2(t,z,i) = 1 - \frac{\Big(\mu(i)-r(i)+\int_{\mathbb{R}}\eta d\nu\Big)\eta(z)}{\sigma^2(t,i) + \int_{\mathbb{R}}\eta^2 d\nu}.
    \end{rcases}
\end{align}
\noindent Hence, we fix these values of $\beta_1$ and $\beta_2$ in the equation \eqref{L24}. We need to analyse equation \eqref{L24} with these particular relevant values of $\beta_1$ and $\beta_2$. A proof of the following theorem on equation \eqref{L24} is presented in Section 4.

\begin{theo} \label{theo45}
Let $K$ be Lipschitz continuous and satisfies (A3). Under (A1) \\
\noi (i) the IPDE \eqref{L24}, \eqref{beta} and \eqref{term} has a unique classical solution $\varphi$ with at most linear growth is $s$;\\
\noi (ii) the partial derivative $\frac{\pa \vf}{\pa s}$ of the solution $\vf$ is bounded.
\end{theo}

\noi On the other hand, we have established in above derivations of this section the parts (i) and (ii) of the following theorem. This theorem is the main result of this paper.

\begin{theo}
Assume (A1) and (A3). Let $\vf$ be the unique classical solution to the IPDE \eqref{L24}, \eqref{beta} and \eqref{term} as asserted in Theorem \ref{theo45}. Then the predictable process $\xi$ as in \eqref{f} is such that $L$ as in \eqref{Lt} is
\begin{itemize}
    \item[i)] a $P$-local martingale,
    \item[ii)] orthogonal to $\bar{M}$, the martingale part of $S^*$,
    \item[iii)] a square integrable martingale.
\end{itemize}
Thus $\varphi(t,S_t,X_t,Y_t)$ is the locally risk minimizing price of a contract with terminal payoff $K(S_T)$ and $\xi$ constitutes the optimal hedging.
\end{theo}
\proof We have already established the parts (i) and (ii) and have obtained
\begin{equation}\label{FS00}
\frac{K(S_T)}{B_T}=\frac{\varphi_T}{B_T}= \varphi_0 + \int_0^T\xi_udS_u^* + L_T
\end{equation}
using \eqref{Lt}, $\vf$ is the classical solution to \eqref{L24}, \eqref{beta} and \eqref{term} with at most linear growth. Due to the square integrability of $S$ and the Lipschitz property of $K$, the left side expression of the above equation is in $L^2(P)$, hence the right side is also in $L^2(P)$. Among three additive terms, the first one $\varphi_0=\varphi(0,S_0,X_0,Y_0)$ is observed hence deterministic. Hence if we show any of the remaining two terms is in $L^2(P)$ then another is also in $L^2(P)$. We note that since $\frac{\pa \vf}{\pa s}$ is bounded continuous function, using the Mean Value Theorem, we can derive for each $t\in [0,T]$ from \eqref{f}
\begin{eqnarray*}
  \mid\xi_t\mid &\le&  \left(\sup_{[0,T]\times \X} \frac{\sigma^2(t,i)}{ \left(\sigma^2(t,i) + \int_{\mathbb{R}}\eta^2(z)\nu(dz)\right)}\right)
\|\frac{\partial\varphi}{\partial s}\|_{\infty} + \left(\sup_{[0,T]\times \X} \frac{\int_{\mathbb{R}} \eta^2(z) \nu(dz)}{ \left(\sigma^2(t,i) + \int_{\mathbb{R}}\eta^2(z)\nu(dz)\right)}\right) \|\frac{\partial\varphi}{\partial s}\|_{\infty} \\
 &<& 2 \|\frac{\partial\varphi}{\partial s}\|_{\infty}.
\end{eqnarray*}
Therefore, $\xi$ is a bounded process adapted to $\calF_{t-}$. Again since $S$ is square integrable, the integral of $\xi$ wrt $S^*$ on $[0,T]$ is in $L^2(P)$. Thus both the facts have been established, namely $\xi$ is an admissible strategy and $L$ is indeed a square integrable martingale. Therefore, \eqref{FS00} is indeed the desired F-S decomposition. Hence the result follows from the discussion in the subsection 3.1.\qed

\section{Option Price Equation}
We begin with a lemma concerning the parameters appearing in the equation \eqref{L24}. This lemma would be used in the subsequent analysis.
\begin{lem}\label{lbeta}
The map $\beta_1: [0,T] \times \mathcal{X} \rightarrow \mathbb{R}$ given by \eqref{beta} is bounded and continuous. The map $\beta_2: [0,T] \times \mathbb{R}\times\mathcal{X} \rightarrow \mathbb{R} $ given by \eqref{beta} is (i) bounded and (ii) continuous in $t$ uniformly in $z$.
\end{lem}
\proof As $\sigma$ is positive valued continuous function on compact set $[0,T]$,  $\inf_{[0,T]\times \mathcal{X} }\sigma(t,i) >0$ for finite $T$. Hence the denominator in the expressions of $\beta_1$ and $\beta_2$ is away from zero. Thus using finiteness of $\nu$ and boundedness of $\eta$, $\beta_1$ and $\beta_2$ are bounded and also continuous in $t$. Again since $\eta$ is bounded, $\sup_{z,i} |\beta_2(t,z,i)- \beta_2(t',z,i)|\to 0$ as $t'\to t$. Hence the proof. \qed

\noindent Consider the evolution problem \eqref{L24} on $\D$ with the terminal condition \eqref{term} and  $\beta_1, \beta_2$ according to \eqref{beta}. A typical expression of $K$, as in the case of call option takes the form $K(s,i,y) = (s-K_1)^+$, where $K_1$ is the strike price. Here $K$ is not differentiable in $s$, thus not in the domain of the operator present in the equation. Hence a classical solution to \cref{L24,term,beta} is not assured. However $K$ belongs to the following space
\begin{equation*}
    V := \left\{\psi : (0,\infty) \times \mathcal{X} \times (0,\infty) \rightarrow \mathbb{R} \textrm{ continuous } \Bigg| \|\psi\|_V := \sup_{s,i,y} \frac{|\psi(s,i,y)|}{1+s} < \infty\right\}.
\end{equation*}
\noindent Clearly $(V,\|\cdotp\|_V)$ is a Banach space consisting of continuous functions with at most linear growth in $s$ bounded in $y$ for fixed $s$ and $i$.

\noi In this section, via an analysis of the mild solution to the evolution problem given in \eqref{L24}, \eqref{term} and \eqref{beta} which is a member of  $C([0,T];V)$, we aim to establish the existence and uniqueness of a classical solution to the evolution problem. To this end, we would first rewrite the evolution problem in a manner, suitable for applying the general theory of abstract evolution problems. Then we establish the existence and uniqueness of the continuous mild solution in Corollary \ref{CauchyP}. For this, we now introduce another SDE on the same probability space
\begin{equation*}
    d\widehat{S}_t = \widehat{S}_t\Big(\big(r(X_t) + \beta_1(t,X_t)\big)dt + \sigma(t,X_t)dW_t\Big),
\end{equation*}
\noindent where $\{W_t\}_{t\geq0}$ is the Brownian motion and $\{X_t\}_{t\geq0}$ is as in (\ref{xdef})-(\ref{ydef}). The above SDE has a strong positive continuous solution whose proof is much simpler than that of Theorem \ref{theoSDE}. Indeed the solution is given by
\begin{equation}\label{shat}
\widehat{S}_t = \widehat{S}_0 \exp\left( \int_0^t\big(r(X_u) + \beta_1(u,X_u)-\frac{1}{2}\sigma(u,X_{u-})^2\big)du+\int_0^t\sigma(u,X_{u-})dW_u\right).
\end{equation}
\noi The solution $\widehat{S} := \{\widehat{S}_t\}_{t\geq0}$ along with $X$ and $Y$ jointly is strongly Markov.  We denote the infinitesimal generator of $\{(\widehat{S}_t, X_t, Y_t)\}_{t\geq0}$ by $\{\widehat{A}_t\}_{t\ge 0}$ which is given by
\begin{equation*}
    \widehat{A}_t\psi(s,i,y) := \Big( \frac{\partial}{\partial y} + \big(r(i)+\beta_1(t,i)\big)s\frac{\partial}{\partial s} + \frac{1}{2}\sigma^2(t,i)s^2\frac{\partial^2}{\partial s^2}\Big)\psi(s,i,y) + \sum_{j\neq i}\lambda_{ij}(y)\big(\psi(s,j,0) - \psi(s,i,y)\big)
\end{equation*}
\noindent where $\psi\in C^\infty_c$. The Feller property of $(\widehat{S},X,Y)$ implies that there exists a continuous \emph{evolution system} (ES) $\{U(u,t)\}_{0\le t \le u\le T}$ on $C^\infty_c$ associated to $\{\widehat{A}_t\}_{t\ge 0}$ (see Definition 5.1.3 \cite{Pazy}). Indeed $U(u,t) \psi(s,i,y)$ is given by $E(\psi(\widehat{S}_u, X_u, Y_u)\mid \widehat{S}_t=s, X_t=i, Y_t=y)$ for every $\psi \in C_c^\infty$. For the interest of our problem, we wish to confirm if  $\{U(u,t)\}_{0\le t\le u\le T}$ is a continuous ES on $V$ associated to $\{\widehat{A}_t\}_{t\ge 0}$. This, along with some differentiability properties of the above ES is presented below whose proofs are given at the end of this section.
\begin{lem}
    \label{lem 4.1} Assume A1(i) and (ii).
    For any $\psi\in V$ and $0 \le t< u \leq T$, the map $\Psi(u,t, \cdot,\cdot, \cdot): \D\to \RR$ given by $\Psi(u,t,s,i,y):= U(u,t)\psi(s,i,y)$ has the following properties.
    \begin{itemize}
    \item [i.] $\Psi$ has at most linear growth in $s$ and is bounded in other variables for each fixed $s$. More precisely,
 $$ \|\Psi(u,t, \cdot,\cdot,\cdot)\|_V \le \|\psi\|_V \left(1+e^{T\sup_{i,t\in [0,T]}\big(r(i) + \beta_1(t,i)\big)}\right) ~~ \textrm{ for all } 0 \le t< u \leq T. $$
    \item[ii.] $\Psi$ satisfies the following integral equation
    \begin{align}
    \nonumber \Psi(u,t,s,i,y) =& \frac{1-F(u-t+y|i)}{1-F(y|i)}\int_0^{\infty}\psi(x,i,y+u-t)\alpha(x;s,i,t, u-t)dx\\
    &+ \int_0^{u-t}\frac{f(y+v|i)}{1-F(y|i)}\sum_{j\neq i} p_{ij}(y+v)\int_0^{\infty}\Psi(u,t+v,x,j,0)\alpha(x;s,i,t, v)dxdv \label{44}
    \end{align}
for all $t\in [0,u), s>0, i \in \mathcal{X}, y>0$ where for every fixed combination of $s$, $i$, $t$, and $v>0$, the map  $x\mapsto\alpha(x;s,i,t,v)$ is the probability density function of the log-normal distribution
$$\mathbf{Lognormal}\left(\ln s+r(i)v+\int_t^{t+v}\big(\beta_1(t',i)-\frac{1}{2}\sigma^2(t',i)\big)dt',~~ \int_t^{t+v}\sigma^2(t',i)dt'\right).$$
\item[iii.]  $\Psi$ is the unique continuous solution to \eqref{44} with values $\Psi(u,t,\cdot,\cdot,\cdot)$ in $V$ for $t\in [0,u)$ and $u\in [0,T]$. Thus $\{U(u,t)\}_{0\le t \le u\le T}$ is a continuous ES on $V$ associated to $\{\widehat{A}_t\}_{t\ge 0}$.
\item[iv.] $\Psi: \bigcup_{0<t<T} \Big( (t,T)\times \{t\}\times (0,\infty)\times \X \times (0,\infty)\Big)\to \RR$ has continuous second order partial derivative w.r.t. $s$ and $\Psi(u, \cdot,s, i, \cdot)$ is in the domain of $D_{t,y}$ where $D_{t,y}\theta(t,y)$ is defined as
    \begin{equation*}
    \lim_{h\rightarrow0}\frac{1}{h}\Big(\theta(t+h,y+h)-\theta(t,y)\Big)
    \end{equation*}
    provided the limit exists. Furthermore, $D_{t,y}\Psi(u,t,s,i,y)$ is a bounded continuous function for each $s>0$.
    \end{itemize}
\end{lem}

\noi We can rewrite \eqref{L24} by substituting the expression of $A_t$ as
\begin{align}\label{IPDE}
    \no &\Big(\frac{\partial}{\partial t} + \frac{\partial}{\partial y} + \big(r(i) + \beta_1(t,i)\big)s\frac{\partial}{\partial s} + \frac{1}{2}\sigma^2(t,i)s^2\frac{\partial^2}{\partial s^2}\Big)\varphi(t,s,i,y) + \sum_{j\neq i}\lambda_{ij}(y)\big(\varphi(t,s,j,0) - \varphi(t,s,i,y)\big) \\
    & + \int_\mathbb{R} \beta_2(t,z,i) \Big(\varphi(t, s(1+\eta(z)), i, y)-\varphi(t, s, i, y)\Big)\nu(dz) = r(i)\varphi(t,s,i,y).
\end{align}

\noindent Hence, using the expression of $\widehat{A}_t$ in the above equation, we have the following evolution problem
    \begin{align}
        \begin{rcases}
       \left(\frac{\partial}{\partial t} + \widehat{A}_t\right)\varphi(t) + B(t)\varphi(t) -R\varphi(t) &= 0\\
        \varphi(T) &= K
        \end{rcases} \label{CP4.1}
    \end{align}
\noindent where for each $t\in [0,T]$, $\vf(t) \in V$, and $\vf(t)(s,i,y)$ is also written as $\vf(t,s,i,y)$; further more, for any $\psi\in V$, $B(t)\psi(s,i,y) := \int_\mathbb{R} \beta_2(t,z,i) \Big(\psi(s(1+\eta(z)), i, y)-\psi(s, i, y)\Big)\nu(dz)$ and $R\psi(s,i,y) := r(i)\psi(s,i,y)$. Using Lemma \ref{lbeta} $B(t)$ is a bounded linear map for each $t\in [0,T]$ and $t\mapsto B(t)$ is continuous. The proof of this fact appears in the Appendix, Lemma \ref{lem B bounded}.

\noindent Let $K$ be in $V$ and $f : [0,T] \times V \rightarrow V$ be a continuous function in $t$ and uniformly Lipschitz continuous on $V$. Theorem 6.1.2 of \cite{Pazy} implies that the initial value problem
\begin{align*}
        \frac{\partial\varphi(t)}{\partial t} = \widehat{A}_t\varphi(t) + f(t,\varphi(t)), \ \ \ \ \ \ \ \
        \varphi(0) = K,
    \end{align*}
\noindent has a unique continuous mild solution which solves another integral equation given by
\begin{equation}
     \nonumber   \varphi(t) = U(t,0)K + \int_0^tU(t,u)f(u,\varphi(u))du, \label{CP4.3}
    \end{equation}
\noindent where $\{U(t,u)\}_{0\le u <t \le T}$ is the ES associated to $\{\widehat{A}_t\}_{t \in [0,T]}$. The obvious counterpart of the above statement for a terminal value problem is stated below.

\begin{prop} \label{theo41}
Under A1(i)-(ii) for every $K$ in $V$, the evolution problem
$$\frac{\partial\varphi (t)}{\partial t}+\widehat{A}_t\varphi(t) + f(t,\varphi(t)) = 0, \ \ \ \varphi(T) = K
$$
has a unique continuous mild solution which solves
\begin{equation}
        \varphi(t) = U(T,t)K + \int_t^T U(u,t)f(u,\varphi(u))du, \label{CP4.4}
\end{equation}
provided $f(t,\vf)$ is continuous in $t$, on $[0,T]$ and uniformly Lipschitz continuous in $\vf \in V$.\\
Here $\{U(t,u)\}_{0\le u <t \le T}$ is the ES associated to $\{\widehat{A}_t\}_{t \in [0,T]}$.
\end{prop}

\noindent It is easy to see that the evolution problem \eqref{CP4.1} is a special case of the above problem where
\begin{equation*}
    f(t,\varphi(t)) =(B(t)-R)\varphi(t),
\end{equation*}
\noindent where $B(t)$ and $R$ are as in \eqref{CP4.1}. Indeed $R$ is a bounded linear operator and $B\in C([0,T];\mathfrak{B}(V))$ where $\mathfrak{B}(V)$ denotes the space of bounded linear operators on $V$ (see Lemma \ref{lem B bounded}), and hence the above $f$ satisfies the conditions of Proposition \ref{theo41}. Thus using Lemma \ref{lem 4.1}(iii) and \eqref{CP4.4}, the following integral equation
\begin{equation}
    \varphi(t) = U(T,t)K + \int_t^T U(u,t)\big(B(u)-R\big)\varphi(u)du \label{gensol}
\end{equation}
\noi has a continuous solution and the solution is the mild solution to \eqref{CP4.1}. We state this as a corollary below.
\begin{coro}
Under A1(i)-(ii) for every $K$ in $V$, the evolution problem \eqref{CP4.1} has a unique continuous mild solution and that solves \eqref{gensol}. \label{CauchyP}
\end{coro}

\noi Having proved this, it remains to establish the regularity of the mild solution to justify that the mild solution is indeed the classical solution. That is immediate for the cases when either $\{\widehat{A}_t\}_t$ are bounded operators or $K$ is in the domain of $\widehat{A}_t$. However, none of these conditions are true in our setting. We can only confirm the applicability of $D_{t,y}$ and $\frac{\pa^2}{\pa s^2}$ on $U(T,t)K$ for each $K\in V$ using Lemma \ref{lem 4.1}(iv). Although $D_{t,y}U(T,t)K(s,i,y)$ has been proved to be a bounded continuous function, $\frac{\pa^2}{\pa s^2} U(T,t)K$ need not be bounded. In particular, the payoff function for call option is $K(s,i,y) = (s-K_1)^+$ and hence $\frac{\pa^2}{\pa s^2} U(T,\cdot)K(s,i,y)$ is unbounded on $(0,T)$ for $s=K_1$. Nevertheless, standard payoff functions including those for put and call options satisfy the following integrability condition.
\begin{itemize}
\item[\textbf{A3.}]
Define $\vp_h(t):= \Delta_h U(T,t)K - \frac{\pa^2}{\pa s^2} U(T,t)K$ where for each $\psi \in V$, and $h>0$, $\Delta_h \psi(s,i,y):= (\psi(s+h,i,y)-2\psi(s,i,y)+\psi(s-h,i,y))/h^2$. Let $K\in V$ such that
$$ \int_{[0,T)} \left\|\frac{\pa^2}{\pa s^2} U(T,t)K \right\|_V dt <\infty, \tab \int_{[0,T)} \left\|\vp_h(t)\right\|_V dt \to 0.$$
Or in other words $U(T,t)K$ is twice partially differentiable uniformly in $L^1([0,T); V)$.
\end{itemize}
\begin{theo} \label{theo5} Assume (A1) and (A3). Let $\varphi$ be a continuous solution to \eqref{gensol}, then, $\varphi(\cdot)(\cdot,\cdot, \cdot)$ in $C(\D)$ is twice differentiable in $s$; and for every $s$, $i$, $\vf(\cdot)(s,i,\cdot)$ is in the domain of $D_{t,y}$ (see Lemma \ref{lem 4.1}(iv) for the definition of $D_{t,y}$.).
    \end{theo}
\proof Let $\vf$ be the continuous solution to the IE \eqref{gensol} which has two additive terms on the right side of the equality sign. The fact that the first term i.e., $U(T,t)K$ is in the domain of $D_{t,y}$ for all $t<T$ follows from a direct application of Lemma \ref{lem 4.1}(iv). Furthermore, Lemma \ref{lem 4.1}(iv) also implies continuity of the partial derivatives. Next we prove that the second term, i.e.,
\begin{equation}\label{right}
\int_t^T U(u,t)\big(B(u)-R\big)\varphi(u)du
\end{equation}
is also in the domain of $D_{t,y}$; and the derivatives are continuous. That would establish the continuous differentiability of the right side expression of \eqref{gensol} and hence of $\vf$, the left side of \eqref{gensol}.

\noi Lemma \ref{lem 4.1}(iv) implies that the integrand of \eqref{right} is in the domain of $D_{t,y}$ and  $u\mapsto D_{t,y} U(u,t)\big(B(u)-R\big)\varphi(u)$ on $(t,T]$ is bounded continuous for each $s$ and $i$. Therefore, by using Mean Value Theorem and then the Dominated Convergence Theorem, we can establish applicability of $D_{t,y}$ on \eqref{right} easily. The argument for continuity of $D_{t,y}$ of \eqref{right} is also straight forward.

\noi Let $g(t):= \frac{\pa^2}{\pa s^2} U(T,t)K \in V$ for all $0\le t<T$. Then from (A3) $g \in L^1([0,T); V)$. Consider the integral equation
\begin{equation}\label{eq1}
w(t) = g(t) +\int_{[t,T)} U(u,t)(B(u) -R)w(u) du, \tab \text{for all } t\in [0,T).
\end{equation}
The proof of existence and uniqueness of a solution to the above equation in the class $L^1([0,T); V)$ is standard and requires direct application of contraction mapping theorem. We further define, $w_h(t):= \Delta_h\vf(t)-w(t)$. Hence a direct calculation gives
$$
w_h(t) = \vp_h(t) + \int_{[t,T)} U(u,t)(B(u) -R)w_h(u) du.
$$
Using Gronwall's inequality we get
\begin{eqnarray*}
  \|w_h(t)\|_V &\le& \|  \vp_h(t)\|_V + \int_{[t,T)} \| \vp_h(r)\|_V \|(B(r) -R)\|_{\mathfrak{B}(V)} \exp ( \int_r^t \|(B(u) -R)\|_{\mathfrak{B}(V)} du )dr \\
   &\le &  \|  \vp_h(t)\|_V  + c\exp (cT) \|  \vp_h(\cdot)\|_{L^1([0,T); V)}
\end{eqnarray*}
where $c=\|(B(\cdot) -R)\|_{C([0,T];\mathfrak{B}(V))}$. Clearly, (A3) implies $\|\vp_h(t)\|_V\to 0$ as $h\to 0$ for all $t<T$. Thus, from above, and (A3) $ \|w_h(t)\|_V \to 0$ as well as $ \|w_h(\cdot)\|_{L^1([0,T); V)} \to 0 $. Therefore, $\vf$, the continuous solution to the IE \eqref{gensol} is twice differentiable in $s$ and the derivative is $w$, the solution to \eqref{eq1} and hence in $L^1$. However, the left side of \eqref{eq1} is continuous in $t$, thus $w$ is also continuous.\qed

\noi {\bf proof of Theorem \ref{theo45}} (i) The IPDE \eqref{L24}, \eqref{beta} and \eqref{term} is rewritten as \eqref{CP4.1}. Thus by using Corollary \ref{CauchyP}, the IPDE  has a unique continuous mild solution $\vf$ in $C([0,T]; V)$ which solves the integral equation \eqref{gensol}. Under (A1) and (A3) Theorem \ref{theo5} implies that this mild solution is in the domain of operators in \eqref{L24}. Hence that mild solution solves \eqref{L24} classically.

\noi (ii) We write down an equation for  $\rho(t,s,i,y):=\frac{\pa \vf}{\pa s}(t,s,i,y)$ by employing partial derivative on both sides of \eqref{IPDE} w.r.t. $s$ variable as below
\begin{eqnarray*}
\lefteqn {\Big(D_{t,y} +\big(r(i) + \beta_1(t,i)\big) + \big(r(i) + \beta_1(t,i)\big)s\frac{\partial}{\partial s} + \sigma^2(t,i)s\frac{\partial}{\partial s} \frac{1}{2}\sigma^2(t,i)s^2\frac{\partial^2}{\partial s^2}\Big)\rho(t,s,i,y)} \\
&&  + \sum_{j\neq i}\lambda_{ij}(y)\big(\rho(t,s,j,0) - \rho(t,s,i,y)\big) \\
&&+ \int_\mathbb{R} \beta_2(t,z,i) \Big((1+\eta(z))\rho(t, s(1+\eta(z)), i, y)-rho(t, s, i, y)\Big)\nu(dz) \\
&&= r(i)\rho(t,s,i,y),
\end{eqnarray*}
or
\begin{align*}
&\Big(D_{t,y} + \big(r(i) + \beta_1(t,i) + \sigma^2(t,i)\big)s\frac{\partial}{\partial s} + \frac{1}{2}\sigma^2(t,i)s^2\frac{\partial^2}{\partial s^2}\Big)\rho(t,s,i,y) + \sum_{j\neq i}\lambda_{ij}(y)\big(\rho(t,s,j,0) - \rho(t,s,i,y)\big) \\
&+ \int_\mathbb{R} \beta_2(t,z,i) \Big((1+\eta(z))\rho(t, s(1+\eta(z)), i, y)-\rho(t, s, i, y)\Big)\nu(dz) \\
&= -\beta_1(t,i) \rho(t,s,i,y).
\end{align*}
and $\rho(T) = K'$  where $K'$ is the almost everywhere derivative of $K$. Now we wish to ensure that $\rho$ indeed solves this classically. To this end we rewrite the above problem in the following way
\begin{align}
        \begin{rcases}
       \frac{\partial}{\partial t}\rho(t) + {\bar{A}}_t\rho(t) + \bar{f}(t)\rho(t) &= 0\\
        \rho(T) &= K'
        \end{rcases} \label{CP4.2}
\end{align}
with an $\bar{f}\in C([0,T];\mathfrak{B}(V))$, whereas $\{\bar{A}_t\}$ is the infinitesimal generator of the solution to another SDE given by
\begin{equation*}
    d\bar{S}_t = \bar{S}_t\Big(\big(r(X_t) + \beta_1(t,X_t)+ \sigma(t,X_t)^2\big)dt + \sigma(t,X_t)dW_t\Big)
\end{equation*}
on the same probability space where $X$ and $W$ are as before. Let $\{\bar U(u,t)\}_{0\le t\le u\le T}$ be given by
$$\bar U(u,t) \psi(s,i,y):= E[\psi(\bar S_u,X_u,Y_u)\mid\bar S_t=s,X_t= i,Y_t=y] .
$$
Then as in Lemma \ref{lem 4.1}, one can establish that $\bar U$ is indeed a continuous ES on $V$. It is easy to see that although $K'\notin V$, $\bar U(u,t)K'$ is well defined and is in $V$ for any $0\le t<u$. Now the successive applications of Proposition \ref{theo41}, Theorem \ref{theo5} and Theorem \ref{theo45}(i) ensure existence of a classical solution to \eqref{CP4.2} which is also asserted to solve an IE
\begin{equation}
\no \rho(t) = \bar{U}(T,t)K' + \int_t^T\bar{U}(u,t)\bar{f}(u)\rho(u)du.
\end{equation}
Endow $\mathcal{V}=L^\infty\big((0,\infty)\times \X\times (0,\infty)\big)$ with essential supremum norm. As $\|\bar{U}(u,t)\rho(t)\|_\mathcal{V} \le \| \rho(t)\|_\mathcal{V}$, and $\bar f$ is also in $C([0,T];\mathfrak{B}(\mathcal{V}))$, we get from above IE
$$\| \rho(t)\|_\mathcal{V} \le \| K'\|_\mathcal{V} + \left(\sup_{t'\in [0,T]} \|\bar{f}(t')\|_{\mathfrak{B}(\mathcal{V})}\right)\int_t^T \|\rho(u)\|_\mathcal{V} du.$$
Using Gronwall's inequality
$$\|\rho(t)\|_\mathcal{V} \le \|K'\|_\mathcal{V} e^{\left(\sup_{t'\in [0,T]} \|\bar{f}(t')\|_{\mathfrak{B}(\mathcal{V})}\right)(T-t)} \le \|K'\|_\mathcal{V} e^{T\left(\sup_{t'\in [0,T]} \|\bar{f}(t')\|_{\mathfrak{B}(\mathcal{V})}\right)}<\infty
$$
for all $t\in [0,T]$. Or in other words, $\frac{\pa \vf}{\pa s}(\cdot,\cdot,\cdot,\cdot)$ is a bounded function. \qed

\noi {\bf proof of Lemma \ref{lem 4.1}} {\bf (i)} The measurability of $\Psi$ follows from its representation as a conditional expectation. Let $\{\mathcal{F}^X_t\}$ denote the filtration generated by the process $X$. Consider $ \widehat{S}$ as in \eqref{shat}. Then
  		\begin{align*}
  		E\left[ \frac{ \widehat{S}_t}{\widehat{S}_0} \bigg|\mathcal{F}^X_t\right] =  E\left[a.e.\lim\limits_{N\to \infty}\prod_{i=1}^{N} \frac{ \widehat{S}_{T_i\wedge t}}{ \widehat{S}_{T_{i-1}\wedge t}} \bigg|\mathcal{F}^X_t\right] \leq\varliminf_{N\to \infty} E\left[\prod_{i=1}^{N} \frac{ \widehat{S}_{T_i\wedge t}}{ \widehat{S}_{T_{i-1}\wedge t}} \bigg|\mathcal{F}^X_t\right],
  		\end{align*}
  		by Fatou's Lemma, where $T_i$ is as in section 2.2 and denotes the $i$th transition time of $X$. Now since for each $i=1,\ldots, N$, $\frac{ \widehat{S}_{T_i\wedge t}}{\widehat{S}_{T_{i-1}\wedge t}}$ are conditionally independent to each other given $\mathcal{F}^X_t$, using \eqref{shat} the expression on the right side of above inequality can be rewritten as
  		$ \varliminf\limits_{N\to \infty} \prod_{i=1}^{N}e^{\int_{T_{i-1}\wedge t}^{T_i\wedge t}\big(r(X_u) + \beta_1(u,X_u)\big)du},$ which is same as $\exp({\int_0^t \big(r(X_u) + \beta_1(u,X_u)\big)du})$, a bounded random variable. Thus using the above observation, we get
\begin{eqnarray*}
|\Psi(u,t,s,i,y)|&\le& E(|\psi(\widehat{S}_u, X_u, Y_u)|\mid \widehat{S}_t=s, X_t=i, Y_t=y)\\
&\le& \|\psi\|_V  E( 1 +\widehat{S}_u\mid \widehat{S}_t=s, X_t=i, Y_t=y)\\
&\le&\|\psi\|_V + \|\psi\|_V E\left[e^{\int_t^u \big(r(X_{t'}) + \beta_1(t',X_{t'})\big)dt'}| X_t=i, Y_t=y\right]s\\
&\le& \|\psi\|_V \left(1+e^{T\sup_{i',t'\in [0,T]}\big(r(i') + \beta_1(t',i')\big)}s\right)
\end{eqnarray*}
for all $0 \le t< u \leq T$, $i \in \mathcal{X}$ and $y \ge 0$. Now using Lemma \ref{lemsup}, the result follows.

\noi {\bf (ii)} Using the functions $F$, $f$, and $n(t)$ as in Proposition \ref{prop}, and tower property of conditional expectation, for all $u> t$
\begin{align}\label{**}
\no    U(u,t)\psi(s,i,y) =&E[\psi(\widehat{S}_u, X_u, Y_u)|\widehat{S}_t=s, X_t=i, Y_t=y]\\
\no =& E\Big[E\left(\psi(\widehat{S}_u, X_u, Y_u) |\widehat{S}_t, X_t, Y_t,T_{n(t)+1}\right)| \widehat{S}_t=s, X_t=i, Y_t=y\Big]\\
= & \int_{v\in (0,\infty)} E\left(\psi(\widehat{S}_u, X_u, Y_u) |\widehat{S}_t=s, X_t=i, Y_t=y,T_{n(t)+1} =t+v\right)\frac{f(y+v|i)}{1-F(y|i)}dv.
\end{align}
\begin{figure}
\centering
\begin{picture}(200,25)(0,-5)
\thicklines
\put(0,0){\line(1,0){200}}
\multiput(0,-5)(50,0){4}{\line(0,1){10}}
\put(0,-10){$0$}
\put(50,-10){$T_{n(t)}$}
\put(70,5){\vector(-1,0){20}}
\put(72,5){$y$}
\put(80,5){\vector(1,0){20}}
\put(100,-10){$t$}
\put(150,-10){$T_{n(t)+1}$}
\put(125,-5){\line(0,1){5}}
\put(125,-10){$u$}
\put(120,5){\vector(-1,0){20}}
\put(130,5){\vector(1,0){20}}
\put(122,5){$v$}
\end{picture}
    \caption{The last and the next transitions of the semi-Markov process $X$}
    \label{time}
\end{figure}
\noi Now for $T_{n(t)+1}>u$ (see Figure \ref{time}), no transition takes place during $[t,u]$. Thus if $Y_t$ is $y$, then $Y_u$, the age at $u$ is equal to $y+u-t$ and the state $X_u$ is identical to $X_t$. Hence using the expression \eqref{shat} of $\widehat{S}$, the conditional distribution of $\widehat{S}_u$ given $\{T_{n(t)+1} >u\}$ and $\mathcal{F}_t$, is lognormal. Thus we get for all $v> u-t$
\begin{align*}
&E\left(\psi(\widehat{S}_u, X_u, Y_u) |\widehat{S}_t=s, X_t=i, Y_t=y,T_{n(t)+1} =t+v\right)\\
&= \int_0^\infty \psi(x,i,y+u-t)\alpha(x;s,i,t,u-t)dx
\end{align*}
where $\alpha$ is as in \eqref{44}. Therefore, by decomposing the domain of $v$ into $(0,u-t]$ and $(u-t,\infty)$ in \eqref{**} and by the tower property of expectation,
\begin{align*}
    U(u,&t)\psi(s,i,y)\\
    =& \int_{u-t}^\infty \frac{f(y+v|i)}{1-F(y|i)}\int_0^\infty \psi(x,i,y+u-t)\alpha(x;s,i,t,u-t)dx dv \\
    & + \int_0^{u-t} \frac{f(y+v|i)}{1-F(y|i)} E\left(E\left[\psi(\widehat{S}_u, X_u, Y_u) | \widehat{S}_{t+v}, X_{t+v}, Y_{t+v} \right]|\widehat{S}_t=s, X_t=i, Y_t=y,T_{n(t)+1} =t+v\right) dv.
\end{align*}
\begin{figure}
\centering
\begin{picture}(200,25)(0,-5)
\thicklines
\put(0,0){\line(1,0){200}}
\multiput(0,-5)(50,0){4}{\line(0,1){10}}
\put(0,-10){$0$}
\put(50,-10){$T_{n(t)}$}
\put(70,5){\vector(-1,0){20}}
\put(80,5){\vector(1,0){20}}
\put(72,5){$y$}
\put(100,-10){$t$}
\put(122,5){$v$}
\put(120,5){\vector(-1,0){20}}
\put(130,5){\vector(1,0){20}}
\put(150,-10){$T_{n(t)+1}$}
\put(185,-10){$u$}
\put(185,-5){\line(0,1){5}}
\end{picture}
    \caption{The last and the next transitions of the semi-Markov process $X$}
    \label{time2}
\end{figure}

\noi See Figure \ref{time2} for the case $v<u-t$. During $[T_{n(t)}, t+v)$, $X$ does not change and the age $Y$ increases monotonically. Finally, $Y_{t+v}=0$ certainly since $T_{n(t) +1} =t+v$ implies that a transition takes place at time $t+v$, or in other words $Y_{t+v}$, the age at $t+v$ is zero. Using the conditional distribution of $X_{t+v}$,  $Y_{t+v}$, and $S_{t+v}$, we get
 \begin{align*}
    U(u,t)\psi(s,i,y)= &\frac{1-F(y+u-t|i)}{1-F(y|i)}\int_0^\infty \psi(x,i,y+u-t)\alpha(x;s,i,t,u-t)dx +\int_0^{u-t}\frac{f(y+v|i)}{1-F(y|i)}\\
    &\sum_{j\neq i} p_{ij}(y+v)
    \int_0^{\infty}E\Big[\psi(\widehat{S}_u,X_u,Y_u)|\widehat{S}_{t+v}=x, X_{t+v}=j, Y_{t+v}=0\Big]\alpha(x; s, i, t, v)dxdv.
\end{align*}
\noi  Thus, $U(\cdot,\cdot)\psi(\cdot,\cdot,\cdot)$ satisfies the following integral equation
    \begin{align}
    \nonumber U(u,t)\psi(s,i,y) =& \frac{1-F(u-t+y|i)}{1-F(y|i)}\int_0^{\infty}\psi(x,i,y+u-t)\alpha(x;s,i,t, u-t)dx\\
\no    &+ \int_0^{u-t}\frac{f(y+v|i)}{1-F(y|i)}\sum_{j\neq i} p_{ij}(y+v)\int_0^{\infty}U(u,t+v)\psi(x,j,0)\alpha(x;s,i,t, v)dxdv.
    \end{align}
\noi Hence (ii) follows.

\noi {\bf (iii)} In (i) and (ii) it is shown that $\Psi$ is a measurable solution to \eqref{44} having  at most linear growth in $s$ and bounded in other variables for each $s>0$. Now for a fixed $t$, and a given $\psi\in V$, due to the property of lognormal density, the right side of \eqref{44} is a continuous function of $s$. The continuity w.r.t. $y$ follows from A1(i)-(ii) and Proposition \ref{prop}(iii)-(iv). Thus $\Psi$, the left side, is also continuous for each $t$ and hence $\Psi(u,t,\cdot,\cdot,\cdot)$ belongs to $V$. Furthermore, in a standard manner, \eqref{44} can be expressed as a fixed point problem of a contraction map on $V$. We recommend the reader to see the proof of Lemma 3.1(i) of \cite{AG2} for the details in a very similar context. Subsequently, a direct application of Banach fixed point theorem completes the proof of (iii).

\noi {\bf (iv)} The proof of this part relies on the results obtained in (i), (ii) and (iii). We use the fact that $\Psi$ satisfies the Volterra integral equation \eqref{44} of second kind. We establish the desired smoothness of $\Psi$ by establishing that for the right side expression of \eqref{44}. On the right hand of \eqref{44} lognormal density $\alpha$ appears in every integral term. We use the smoothness, finite second moment property and estimates of partial derivatives of $\alpha$ for establishing desired smoothness of the integral terms. The details of the proof is presented in the following three steps.

\noi \textbf{Step 1.} In this step we would check the applicability of $D_{t,y}$ on the first additive term on the right of \eqref{44}. Proposition \ref{prop}(iii) asserts that under A1(i), $F$ is twice differentiable. Therefore it is enough to verify that the domain of $D_{t,y}$ contains
\begin{equation*}
\int_0^{\infty}\psi(x,i,y+u-t)\alpha(x;s,i,t,u-t)dx.
\end{equation*}
It is important to note that $\psi$ in the integrand is merely continuous and thus need not be differentiable in $t$ or $y$. However, the image of $D_{t,y}$ on the above function is the limit of
\begin{equation*}
\frac{1}{\vp}\Big[\int_0^{\infty}\psi\big(x,i,y+u-t\big)\big(\alpha(x;s,i,t+\vp, u-t-\vp)-\alpha(x;s,i,t, u-t)\big)\Big]dx
\end{equation*}
as $\vp$ tends to zero, provided that the limit exists. Due to the continuous differentiability of the p.d.f. $\alpha(x;s,i,t,v)$, w.r.t. $t$ and $v$, the above expression can be rewritten using Mean Value Theorem as
\begin{equation}
\int_0^{\infty}\psi\big(x,i,y+u-t\big)\left(\frac{\partial}{\partial t} -\frac{\partial}{\partial v}\right) \alpha \big(x;s,i,t+\vp_1, u-t-\vp_1) dx \label{42}
\end{equation}
for some $0<\vp_1(x,s,i,t,u)<\vp$. Again a direct calculation shows that both $\frac{\partial\alpha}{\partial t}$ and $\frac{\partial\alpha}{\partial v}$ are of the form $\alpha(x;s,i,t,v)O(\ln^2|x|)$. It is important to note that due to the presence of $\vp_1$ at the last two arguments of $\frac{\partial\alpha}{\partial t}$ and $\frac{\partial\alpha}{\partial v}$, those arguments also depend on the $x$ variable. However, $\vp_1\in (0,\vp)$, and for any given $u>0$ and $t$ in $(0,u)$, $\vp$ can be chosen so that $0<\vp<u-t$. Therefore, due to the monotonicity of the tail decay of lognormal density w.r.t. the parameter values, there is an $x'$, large enough and some $t'\in [t,t+\vp]$, $v'\in [u-t-\vp,u-t]$ such that
$$\sup_{\vp_1\in (0,\vp)} \alpha(x;s,i,t+\vp_1,u-t-\vp_1) = \alpha(x;s,i,t',v')$$
for all $x\ge x'$. On the other hand, $\sup_{\vp_1\in (0,\vp)} \alpha(\cdot;s,i,t+\vp_1,u-t-\vp_1)$ is bounded on $[0,x']$. Hence we first use the estimates of $\frac{\pa \alpha}{\pa t}$ and $\frac{\pa \alpha}{\pa v}$ in \eqref{42} and subsequently write the integral \eqref{42} as sum of two integrals by decomposing the domain $(0,\infty)$ as union of $(0,x')$ and $[x', \infty)$. For the first integral with a finite domain and uniformly bounded integrand, the convergence is obvious due to the dominated convergence theorem. Now we consider the remaining part. As $\psi$ is of at most linear growth with respect to $x$, there exists a positive constant $c_1$ such that the absolute value of the integrand in \eqref{42} is dominated by $c_1(1+x)(1+\ln^2(|x|))\alpha(x;s,i,t',v')$ on the interval $[x',\infty)$ for some $t'\in [t,t+\vp]$, $v'\in [u-t-\vp,u-t]$. The integral of this with respect to $x$ over $[x',\infty)$ is finite. The finiteness is immediate, since $\alpha$ is the p.d.f. of a random variable with finite variance and since there is a sufficiently large $c_2$ such that $(1+x)(1+\ln^2|x|)\leq c_2(1+x^2)$, for all $x\ge 0$.
Thus, if
\begin{equation}
\lim_{(t',v')\to (t,v)}\int_{x'}^{\infty} (1+x^2)\alpha(x;s,i,t',v') dx = \int_{x'}^{\infty} \lim_{(t',v')\to (t,v)} (1+x^2)\alpha(x;s,i,t',v') dx <\infty,\label{45}
\end{equation}
using the General Lebesgue Convergence Theorem (Theorem 4.17 \cite{RO}), and the assertion of convergence of integrals on $(0,x')$, we can conclude that  \eqref{42} converges to
\begin{equation}
\int_{0}^{\infty}\psi\big(x,i,y+u-t\big)\left(\frac{\partial}{\partial t} - \frac{\partial}{\partial v}\right)\alpha(x;s,i,t,u-t)dx \label{42a}
\end{equation}
as $\vp$ tends to zero. The equality \eqref{45} can be justified by applying Vitali's convergence theorem. To this end, we note that the integrand on the left side of \eqref{45} is $(1+x^2)\alpha(x;s,i,t',v')$, a product of a quadratic function and a log normal density. This is a uniformly integrable family of functions in $x$ with family-parameters $(t', v')$ which vary on a bounded set away from $\RR\times\{0\}$. This family is also tight as a consequence of tightness of Gaussian measures with bounded means and variances. Therefore Vitali's convergence theorem is applicable in establishing \eqref{45}. Hence, we conclude that \eqref{42a} is the limit of \eqref{42}. Using a very similar argument as above, one can prove the continuity of \eqref{42a} also. We omit the details. Thus, first additive term on the right of \eqref{44} is in the domain of $D_{t,y}$ and the image of $D_{t,y}$ is also continuous.

\noi \textbf{Step 2.} The second term on the right of \eqref{44}, to be denoted by $\gamma$ now onward, is more involved than the first term. This term is a double integral with one of the limits depending on $t$ variable. The variable $t$ appears in the argument of continuous function $\Psi$, but not in the form of $t-y$. We check if this term is in the domain of $D_{t,y}$. The analysis is somewhat similar to the treatment of the first term. Here $D_{t,y}\gamma$ is the limit of the following expression
\begin{align*}
    &\frac{1}{\vp}\Big[\int_0^{u-t-\vp}\frac{f(y+v+\vp|i)}{1-F(y+\vp|i)}\sum_{j\neq i} p_{ij}(y+v+\vp)\int_0^{\infty}\Psi(u,t+v+\vp,x,j,0)\alpha(x;s,i,t+\vp,v)dx dv\\
    &- \int_0^{u-t}\frac{f(y+v|i)}{1-F(y|i)}\sum_{j\neq i} p_{ij}(y+v)\int_0^{\infty}\Psi(u,t+v,x,j,0)\alpha(x;s,i,t,v)dxdv\Big]
\end{align*}
provided the limit exists. After a suitable substitution, the above expression can be rewritten as
\begin{align}
    \nonumber &
     \int_\vp^{u-t}\sum_{j\neq i}p_{ij}(y+v)\int_0^{\infty}\Psi(u,t+v,x,j,0)\beta_{\vp}(x,s,i,t,v,y)dxdv\\
    &-\frac{1}{\vp}\int_0^{\vp}\frac{f(y+v|i)}{1-F(y|i)}\sum_{j\neq i} p_{ij}(y+v)\int_0^{\infty}\Psi(u,t+v,x,j,0) \alpha(x;s,i,t,v)dxdv \label{43}
\end{align}
     where $\beta_{\vp}(x,s,i,t,v,y)=\dfrac{1}{\vp}\Big[\dfrac{f(y+v|i)}{1-F(y+\vp|i)}\alpha(x;s,i,t+\vp,v-\vp)-\dfrac{f(y+v|i)}{1-F(y|i)}\alpha(x;s,i,t,v)\Big]$.
The expression in \eqref{43} has two additive terms. For showing convergence of first term involving repeated integrals, we observe the following. Since $f$ and $\alpha$ are continuously differentiable, using Mean value theorem we can rewrite
\begin{align*}
     \beta_{\vp} =& f(y+v|i) \Big(\dfrac{\partial}{\partial t} + \dfrac{\partial}{\partial y} -\dfrac{\partial}{\partial v}\Big)\dfrac{\alpha(x;s,i,t+\vp_1,v-\vp_1)}{1-F(y+\vp_1|i)}
\end{align*}
\noi for some, $0<\vp_1<\vp$. Due to Proposition \ref{prop}(v), $f$ is bounded and $\frac{1}{1-F}$ is bounded on compact. Thus the inner integral of the first term looks similar to \eqref{42} with one major difference that there $\psi$ is in $V$ but here $\Psi$ is a $V$ valued function of $t$. However, in the part (i), it is established that $ \Psi(u,t, \cdot, \cdot, \cdot)$ is dominated by a fixed member in $V$ uniformly in $t$. Therefore, one can mimic the arguments presented in step 1, to prove that $\lim_{\vp\to 0}\int_0^{\infty} \Psi(u,t+v,x,j,0) \beta_{\vp} (x,s,i,t,v,y)dx $ exists and is equal to $\int_0^{\infty}\Psi(u,t+v,x,j,0)\beta_{0+}(x,s,i,t,v,y)dx$. Since the outer integral is on a bounded domain, to assure the convergence of this term, using the estimates of partial derivatives of $\alpha$, it would be sufficient if we have $\sup_{t'\in (t,t+\vp), v'\in (0,u-t)}\int_0^{\infty}(1+x^2)\alpha(x;s,i,t',v')dx$ is finite.
This follows as the density functions are corresponding to distributions with variances lying on a bounded set. Indeed, for a fixed $s$ and $i$ this supremum is less than or equal to
$$1+ s^2 (\exp( T \sup_{i,t'\in [0,T]}\sigma^2(t',i))-1) \exp(2T\sup_{i,t'\in [0,T]}\big(r(i) + \beta_1(t',i)\big)).$$

\noi Now we address the convergence issue of the second term in \eqref{43}. Clearly \eqref{45} implies continuity of the map
\begin{equation*}
v\mapsto\int_0^{\infty}\Psi(u,t+v,x,j,0)\alpha(x,s,i,t,v)dx
\end{equation*}
on $v>0$ and existence of finite right limit at $v=0$. This integral is multiplied by a continuous function in $v$.
Hence the second term converges to $ -\frac{f(y|i)}{1-F(y|i)}\sum_{j\neq i} p_{ij}(y)\Psi(u,t,s,j,0)$ as $\vp$ goes to zero. Thus, $\gamma$ is in the domain of $D_{t,y}$ and hence from Steps 1 and 2, the left of Equation \eqref{44} is in the domain of $D_{t,y}$. Or in other words, $\Psi$ is in the domain of $D_{t,y}$. Finally from the above derivations, $D_{t,y}\Psi(u,t,s,i,y)$ is equal to
\begin{eqnarray*}
\frac{1-F(u-t+y|i)}{(1-F(y|i))^2}f(y |i)\int_0^{\infty}\psi(x,i,y+u-t)\alpha(x;s,i,t, u-t)dx\\
+ \frac{1-F(u-t+y|i)}{1-F(y|i)}\int_0^{\infty}\psi(x,i,y+u-t)\Big(\frac{\partial}{\partial t} -\dfrac{\partial}{\partial v}\Big)\alpha(x;s,i,t, u-t)dx\\
+\int_0^{u-t}\sum_{j\neq i}p_{ij}(y+v)f(y+v|i) \int_0^{\infty}\Psi(u,t+v,x,j,0) \Big(\dfrac{\partial}{\partial t} + \dfrac{\partial}{\partial y} -\dfrac{\partial}{\partial v}\Big)\dfrac{\alpha(x;s,i,t,v)}{1-F(y|i)} dxdv\\
-\frac{f(y|i)}{1-F(y|i)}\sum_{j\neq i} p_{ij}(y)\Psi(u,t,s,j,0).
\end{eqnarray*}
The continuity of this function w.r.t. $u, t, s$ and $y$ can easily be obtained using the estimates of partial derivatives of $\alpha$ as before. However, we omit the details. Furthermore, from above expression one can easily figure out that $D_{t,y} \Psi(\cdot,\cdot,s,\cdot,\cdot)$ is bounded on $\bigcup_{0<t<T} \Big( (t,T)\times \{t\}\times \X \times (0,\infty)\Big)$ for each fixed $s$.

\noi \textbf{Step 3.}
\noi In this step we would check the second order partial differentiability of $\Psi$ w.r.t. $s$ variable as well as the continuity of the derivative. We recall that $\alpha$ is twice differentiable w.r.t. $s$ and both $\psi$ and $\Psi(u,t,\cdot,\cdot,\cdot)$ are in $V$ (continuous and at most of linear growth in first variable). Further more, $\frac{\partial\alpha}{\partial s}(x;s,i,t,v)=\frac{1}{s}O(\ln|x|)\alpha(x;s,i,t,v)$. Thus by mimicking the earlier steps, for checking partial differentiability of right side of \eqref{44} w.r.t. $s$, we need to consider the following family of dominating integrands $(v,x)\mapsto \frac{1}{s+\vp_1}|x|^2\alpha(x;s+\vp_1,i,t,v)$ with family parameter $\vp_1$. In view of the General Lebesgue Convergence Theorem (Theorem 4.17 \cite{RO}) we need to investigate its integral's convergence as $\vp$ tends to zero. We appeal to Vitali's convergence theorem in this connection. To this end we notice that   $\frac{1}{s+\vp}|x|^2\alpha(x;s+\vp,i,t,v)$ is  uniformly integrable family of functions in $x$ with family-parameter $\vp\ll 1$. This family is also tight, as $\vp$ is taken from a bounded set. Indeed, for our case it is enough to consider $|\vp|< s/2 $. Next, we also notice that
\begin{equation*}
v\longmapsto\int_0^{\infty}\frac{1}{s+\vp}|x|^2\alpha(x;s+\vp,i,t,v)dx,
\end{equation*}
\noi is a bounded function for fixed $s,t$ and also a member of a uniformly bounded family for $\vp\ll 1$. Since the integration w.r.t. $v$ is on a finite range, the dominated convergence theorem can be applied at this stage so as to pass the limit and finally obtain the partial differentiability of right side of \eqref{44} w.r.t. $s$. In a similar manner existence of partial derivative with respect to $s$ of any higher order can be shown successively.     \qed

\section{Conclusion}	
In this paper we have considered regime switching extension of the geometric L\'{e}vy process for asset price modelling. Although the Markov regimes are more common in the literature, we have considered the age dependent semi-Markov process which gives a rather general type of regime switching. Under this model assumption we derive a theoretical fair price of a class of European style contracts using the locally risk minimizing approach. The price function is shown to satisfy an IPDE \eqref{IPDE}. The existence and uniqueness of a classical solution to the equation is also established. We have also found out an expression of the optimal hedging in \eqref{f}. Or in other words, if $\xi(t,s,i,y)$ denotes the number of units invested in the risky asset with immediate (left limit) price $s$ at time $t$ when the immediate market regime is at $i$ with age $y$ then
$$ \xi(t,s,i,y)= \frac{\sigma(t,i)^2\frac{\partial\varphi}{\partial s} (t,s,i,y) + \frac{1}{s} \int_{\mathbb{R}} \Big(\varphi(t, s(1+\eta(z)), i, y) - \varphi(t,s,i,y) \Big)\eta(z) \nu(dz)}{ \left(\sigma(t,i)^2  + \int_{\mathbb{R}}\eta^2(z)\nu(dz)\right)}.
$$
We have also obtained an alternative way to write the option price function using integral equations. This equation appears at \eqref{gensol}. In this the evolution operators appear which can again be calculated by solving another set of integral equations as in \eqref{44}. Thus it is possible to write the option price function as a solution of a system of integral equations. This observation might lead instead of a finite difference method to an alternative numerical scheme involving quadrature method for finding the option price. A systematic critical comparison of these methods might be an interesting research direction. The computation of hedging strategy which involves a partial derivative and an integration of price function can also be discussed under both of the numerical approaches.

\noi There could be many other subsequent studies depending on these results by extending the asset price model. Local volatility extension is of course one possibility. The component-wise semi-Markov regime (CSM) could be another. In the CSM setting, each of the regime dependent market parameters namely, $\mu$, $\sigma$ and $r$ is allowed to be driven by a separate semi-Markov process. These processes could be independent or correlated.

\noi It is interesting to note that the expression of the option price function and that of the hedging function involve the drift parameter $\mu$. However, the estimation of $\mu$ is rather tricky when the dynamics has regime switching and the volatility is not negligibly small. Therefore, a relevant filtering problem should be able to find an application to the derivative pricing problem in the present context.

\appendix
\section[A]{Appendix}
\textbf{Proof of Theorem \ref{theoSDE}:}
If \eqref{E23} has a solution $\{S_t\}$, the jumps of this process originates from the last term on the RHS of \eqref{E23}. So we can write
\begin{equation}
    \Delta S_t = S_t - S_{t-} = S_{t-}\int_{\mathbb{R}} \eta(z)N(dz, \{t\}). \label{a2}
\end{equation}
Thus for any finite measurable function $f$,
\begin{equation}\label{a2a}
f\left(\frac{\Delta S_t}{S_{t-}}\right) = \int_{\mathbb{R}} f(\eta(z))N(dz, \{t\}).
\end{equation}

\noindent From \eqref{E23}, we have
\begin{equation}
    dS_t^c = S_{t-}\Big(\mu_{t-}dt+\sigma_{t-}dW_t\Big) \label{a3}
\end{equation}
since only the first two terms on RHS of \eqref{E23} contributes to the continuous part. Hence
\begin{equation}
    d[S]_t^c = S_{t-}^2 \sigma_{t-}^2 dt. \label{a4}
\end{equation}
Let $\tau := \text{min} \{t>0 : S_t \leq 0\}$ is a stopping time and let $Z_t = \ln S_t$ for $t<\tau$. Applying It\^{o}'s formula on $\ln S_t$ for
$0 \leq t < \tau$ and using \eqref{a2}-\eqref{a4}, we get
\begin{align*}
   dZ_t =& \ \frac{dS_t^c}{S_{t-}} - \frac{1}{2} \frac{d[S^c]_t}{S_{t-}^2} + \ln S_t - \ln S_{t-}\\
   =& \ \mu_{t-}dt + \sigma_{t-}dW_t -\frac{1}{2} \sigma_{t-}^2 dt + \ln ( 1 + \frac{\Delta S_t}{S_{t-}}).
\end{align*}
Using \eqref{a2a}, we get,
\begin{equation*}
   dZ_t = (\mu_{t-} -\frac{1}{2} \sigma_{t-}^2) dt + \sigma_{t-}dW_t  + \int_{\mathbb{R}} \ln (1 + \eta(z))N(dz, \{t\}).
\end{equation*}
By integrating from $0$ to $t$, where $0 \leq t < \tau$,
\begin{align*}
    \frac{S_t}{S_0} =& \exp(Z_t-Z_0)\\
    =& \ \exp\left( \int_0^t(\mu_{u-} -\frac{1}{2} \sigma_{u-}^2) du + \int_0^t \sigma_{u-}dW_u  + \int_0^t \int_{\mathbb{R}} \ln (1 + \eta(z))N(dz, du)\right).
\end{align*}

\noindent Hence the solution to SDE \eqref{E23} has the above form for $0 \leq t < \tau$. Due to the finiteness of $\nu$ and the lower bound of $\eta$, namely $\eta(z) >-1$ for all $z$ in $\RR$, $\int_0^t \int_{\mathbb{R}} \ln (1 + \eta(z))N(dz, du)$ is finite for all $t$ and almost every $\omega$.
If possible, choose $\omega \in \Omega$ such that $\tau(\omega)$ is finite.
Thus by letting $ t \uparrow \tau(\omega)$ in the above expression of $S$, $S_{\tau(\omega)-} > 0$. Hence non-positivity may occur only by a jump.
Equation \eqref{a2} makes it clear that non-positivity of $S_t$ does not happen due to a jump with our assumption $\eta(Z_1) > -1$. Hence $\tau = \infty$ $P$ a.s. and $S_t > 0$ $P$ a.s. for all $t\in (0, \infty)$.

\noi The above analysis implies that if solution exists it is unique and positive. The existence is straight forward since a direct calculation would show that the above mentioned expression of $S$ satisfies the SDE.
\noi For establishing the square integrability, we note the following. Clearly $S_t$ can be written as a product of a conditionally log-normal random variable and the term $\exp(\int_0^t \int_{\mathbb{R}} \ln (1 + \eta(z))N(dz, du))$ where both are independent. We further note that the log-normal random variable has bounded parameters on $[0,T]$. Therefore it is sufficient to check if
$E\left[\exp(2\int_0^t \int_{\mathbb{R}} \ln (1 + \eta(z))N(dz, du))\right]$ is bounded on $[0, T]$.

\noi We first note that $|N_t|:=N(\mathbb{R}\times[0,t])$ is finite a.s. as $|\nu|<\infty$. Therefore
\begin{align*}\label{inteta1}
E\left[ \exp\left(2\int_{0}^{t  }\int_{\mathbb{R}}\ln(1+ \eta(z))\,N(dz, du)\right)\right]&=E\left[\prod_{i=1}^{|N_t| }(1+ \eta(z_i))^2\right]\\
&=E\left[E\left[\prod_{i=1}^{|N_t| }(1+ \eta(z_i))^2\bigg||N_t|\bigg|\right]\right]
\end{align*}
where $\{(z_i,t_i)\mid i=1,\ldots, |N_t|\}$ are the point masses of $N$ on $\mathbb{R}\times[0,t]$. Since $(1+\eta(z_1)),\ldots,(1+\eta(z_{|N_t|}))$ are conditionally independent and identically distributed given $|N_t|=n$, the right side is equal to
$$\sum_{n=1}^{\infty}[E(1+ \eta(z_1))^2]^nP(|N_t|=n) = \sum_{n=1}^{\infty}(1+c)^ne^{-t|\nu|}\frac{(t|\nu|)^n}{n!} =e^{-t|\nu|}\exp\left(t|\nu|(1+c)\right)
= \exp\left(ct|\nu|\right)$$
which is clearly bounded on $[0,T]$, where $E\left[(1+ \eta(z_1))^2-1 \right]=c<\infty$ as $\eta$ is bounded.
\qed

\begin{lem}
\label{lemsup}
\noi For any nonnegative $c$,
\begin{equation*}
    \sup_{s\in(0,\infty)}\Big(\frac{1+cs}{1+s}\Big) \leq 1+c.
\end{equation*}
\end{lem}
\proof
\noi We write $(0,\infty) = (0,1] \cup (1,\infty).$ We check for supremum over $(0,1]$ and $(1,\infty)$ separately. First we note that
\begin{equation}
    \sup_{s\in(0,1]} \frac{1+cs}{1+s} \leq \frac{1+c}{1} = 1+c.
\end{equation}

\noindent Again, since, $0<\frac{1}{s}<1$ for $s\in(1,\infty)$, we have
\begin{align*}
    \sup_{s\in(1,\infty)}\Big(\frac{1+cs}{1+s}\Big) &= \sup_{s\in(1,\infty)}\Big(\frac{\frac{1}{s}+c}{\frac{1}{s}+1}\Big)
    \leq \frac{1+c}{0+1} = 1+c.
\end{align*}
\noi Thus
\begin{align*}
\sup_{s\in(0,\infty)}\Big(\frac{1+cs}{1+s}\Big) &= \max\Big(\sup_{s\in(0, 1)}\Big(\frac{1+cs}{1+s}\Big), \sup_{s\in(1,\infty)}\Big(\frac{1+cs}{1+s}\Big)\Big)\leq 1+c.
\end{align*} \qed

\begin{lem} Let $\beta_2: [0,T]\times \RR \times \mathcal{X} \to \RR$ be a bounded continuous function, such that $\beta_2(t,z,i)$ is continuous in $t$ uniformly in $z$ then $B:[0,T]\to \mathfrak{B}(V)$ as in \eqref{CP4.1} is continuous, where $\mathfrak{B}(V)$ is the space of bounded linear maps from $V$ to $V$ with subordinate norm.
\label{lem B bounded}
\end{lem}
\proof We define $\overline{\beta_2} := \sup_{(t,z,i)}|\beta_2(t,z,i)|$.
Using Lemma \ref{lemsup} with $c= 1 + \eta(z) > 0$ for each $z$,
\begin{align*}
\|B(t) \psi \|_{V} = &
        \sup_{(s,i,y)\in(0, \infty)\times \mathcal{X} \times (0,\infty)}\left|
        \int_{\mathbb{R}}\beta_2(t,z,i)\frac{\psi(s(1+\eta(z)),i,y)-\psi(s,i,y)}{1+s} \nu (dz)\right|\\
\leq & \  \overline{\beta_2} \sup_{(s,i,y)}
        \Big[\int_{\mathbb{R}}\frac{1+s(1+\eta(z))}{1+s}
        \left|\frac{\psi(s(1+\eta(z)),i,y)}{1+s(1+\eta(z))}\right| + \left|\frac{\psi(s,i,y)}{1+s}\right| \nu(dz)\Big]\\
         \leq & \ \overline{\beta_2}\Big[\int_{\mathbb{R}}\left(( 2 + \eta(z)) \| \psi \|_V + \| \psi \|_V\right)\nu (dz)\Big]\\
        = & \ \overline{\beta_2} \| \psi \|_V \Big[\int_\mathbb{R}(3 + \eta(z))\nu (dz)\Big]\\
        = & \ \overline{\beta_2} \| \psi \|_V \Big ( 3 \nu \mathbb{(R)} + \int \eta d\nu\Big) < \infty,
\end{align*}

\noi since $\nu$ is a finite measure and $\eta$ is a bounded function. Hence $\| B(t) \|_{\mathfrak{B}(V)} \leq \overline{\beta_2} \Big( 3 \nu{\mathbb{(R)}}+ \int \eta d\nu\Big)$. Thus $B(t)$ is a bounded linear map for each $t\in [0,T]$.

\noi Again since $\beta_2(t,z,i)$ is continuous in $t$ uniformly in $z$, in the similar manner as above, for nonzero $\psi$
$$\frac{\|B(t) \psi -B(t') \psi \|_{V} }{\| \psi \|_V} \le \sup_{z,i}|\beta_2(t,z,i)- \beta_2(t',z,i)|\Big ( 3 \nu \mathbb{(R)} + \int \eta d\nu\Big) \rightarrow 0
$$
as $t'$ tends to $t$.
Thus  $\lim_{t'\to t}\| B(t)-B(t') \|_{\mathfrak{B}(V)} \rightarrow 0$. Hence the result.
\qed


\begin{thebibliography}{27}

\bibitem{AA} Aase, K. K., Contingent claims valuation when the security price is a combination of an Itô process and a random point process, Stochastic Process. Appl., 28, 185-220, 1988.


\bibitem{BC} Bardhan, I. and Chao, X., Pricing options on securities with discontinuous returns, Stochastic Process. Appl., 48, 123-137, 1993.

\bibitem{BS}
    Black, F.; Scholes, M., The pricing of options and corporate liability, Journal of Political Economy, 81, 637-659, 1973.

\bibitem{BGG}
    Basak, Gopal K.; Ghosh, Mrinal, K. and Goswami, A., Risk minimizing option pricing for a class of exotic options in a Markov-modulated market, Stoch. Ann. App., 29, 259-281, 2011.

\bibitem{BGO}
    Biswas, Arunangshu; Goswami, A.; Overbeck, Ludger, Option Pricing in a Regime Switching Stochastic Volatility Model, Statistics \& Probability Letters, 138, 116-126, 2018.

\bibitem{Bulla}
    Bulla, Jan and Bulla, Ingo, Stylized facts of financial time series and hidden semi-Markov models, Computational Statistics and Data Analysis, 51, 2192-2209, 2006.

\bibitem{Chang}
    Chang-Jin, Kim and Charles, R. Nelson, Business cycle turning points, A new Coincident Index, and tests of Duration Dependence based on a Dynamic Factor Model with Regime Switching, Review of Economics and Statistics, 80(2), 188-201, 1998.

\bibitem{RC}
  Cont, Rama, Financial modelling with jump processes, Chapman \& Hall/CRC, 2004.

\bibitem{CH} Chan, Terence, Pricing contingent claims on stocks driven by L\'{e}vy processes, Ann. Appl. Probab., 9(2), 504-528, 1999.

\bibitem{TP}
    Das, M. K.; Goswami, A. and Patankar, T., Pricing derivatives in a regime switching market with time inhomogeneous volatility, Stoch. Anal. Appl., 36(4), 700-725, 2018.

\bibitem{DA}
    Das, M. K. and Goswami, A.,
Testing of binary regime switching models using squeeze duration, Int. J. Financ. Eng., 6(1), 1950006, 20 pp, 2019.

\bibitem{DG}
    Deshpande, A. and Ghosh, M.K, Risk minimizing option pricing in a regime switching market, Stoch. Ann. App., 26(2), 313-324, 2008.

\bibitem{DKR}
    DiMasi, G. B.; Kabanov, Y. and Runggaldier, W.J, Mean-variance hedging of options on stocks with Markov volatility, Theory Probab. Appl., 39, 173-181, 1994.

\bibitem{Elliot2005}
    Elliott, R. J.; Chan, L. L. and Siu, T. K., Option pricing and Esscher transform under regime switching, Annals of Finance, 1, 423-432, 2005.

\bibitem{Elliot2007}
    Elliott, R. J.; Siu, T. K.; Chan, L. L. and Lau, J. W., Pricing options under a generalized Markov-modulated jump-diffusion model, Stochastic Analysis and Applications, 25, 821-843, 2007.

\bibitem{FSond}
    F\"{o}llmer, H. and Sonderman, D., Hedging of non-redundant contingent claims, in Contribution to Mathematical Economics, 205-223, 1986.

\bibitem{FS}
    F\"{o}llmer, H. and Schweizer, M., Hedging of contingent claims under incomplete information, Applied Stochastic Analysis, Stochastics Monographs, 5, 389-414, 1991.

\bibitem{AG1}
    Ghosh, M. K. and Goswami, A., Risk minimizing option pricing in a semi-Markov modulated market, SIAM J. Control Optim., 48, 1519-1541, 2009.

\bibitem{AG2}
  Goswami, A.; Patel, J. and Shevgaonkar, P., A system of non-parabolic PDE and application to option pricing, Stoch. Anal. Appl., 34(5), 893-905, 2016.

\bibitem{GN}
    Goswami, A. and Nandan, S., Convergence of estimated option price in a regime switching market, Indian J. Pure Appl. Math., 47(2), 169-182, 2016.

\bibitem{Guo}
    Guo, X., Information and option pricings. Quantitative Finance, 1, 38-44, 2001.

\bibitem{Pazy}
    Pazy, Amnon., Semigroups of linear operators and applications to partial differential equations., 44. Springer Science \& Business Media, 1983.

\bibitem{PR}
    Protter, Philip and Shimbo, K., No arbitrage and general semimartingales, Markov processes and related topics: a Festschrift for Thomas G. Kurtz. Institute of Mathematical Statistics, 267-283, 2008.

\bibitem{RO} Royden, H. L., Real analysis. Third edition. Macmillan Publishing Company, New York, 1988.

\bibitem{Sch1990}
    Schweizer, M., Risk-minimality and orthogonality of martingales, Stoch. Process. Appl. 30, 123-131, 1990.

\bibitem{Sch1992}
    Schweizer, M., Martingale densities for general asset prices, J. Math. Econ., 21, 363-378, 1992.

\bibitem{Sch1993}
    Schweizer, M., Option hedging for semi-martingales, Stochastic Processes and their Applications, 37, 339-363, 1993.

\bibitem{Sch2001}
    Schweizer, M., A Guided Tour through Quadratic Hedging Approaches, E. Jouini, J. Cvitani\`{c}, M. Musiela (eds.), Option Pricing Interest Rates and Risk Management, Cambridge University Press, 538-574, 2001.

\bibitem{Shir}
    Shiryaev Albert, Essentials of Stochastic Finance, World Scientific, 1999.

\bibitem{Siu}
    Siu, T. K.; Yang, H. L. and Lau, J. W., Pricing currency options under two-factor Markov-modulated stochastic volatility models, Insurance: Mathematics and Economics, 43, 295-302, 2008.

\bibitem{Siu2}Siu, T. K., A hidden Markov-modulated jump diﬀusion model for European option pricing, Hidden Markov models in finance, in Chapter Internat. Ser. Oper. Res. Management Sci., 209, 185-209, Springer, New York, 2014.

\bibitem{Su}
    Su, Xiaonan; Wang, Wensheng and Hwang, Kyo-Shin, Risk-minimizing option pricing under a Markov-modulated jump-diffusion model with stochastic volatility, Statistics \& Probability Letters, 82, 1777-1785, 2012.

\bibitem{VV} Vandaele, Nele; Vanmaele, Mich\'{e}le, A locally risk minimizing hedging strategy for unit-linked life insurance contracts in a L\'{e}vy process financial market, Insurance: Mathematics and Economics, 42, 1128-1137, 2008.
\end{thebibliography}
\end{document}